\documentclass[useAMS,usenatbib]{mn2e}

\usepackage{subfigure}
\usepackage{amsmath}	
\usepackage{amssymb}	
\usepackage{graphicx}
\usepackage{algorithm}
\usepackage[noend]{algpseudocode}
\usepackage{mathrsfs}
\usepackage{url}

\newcommand{\bz}{\bmath{z}}

\newcommand{\bA}{\bmath{A}}
\newcommand{\bB}{\bmath{B}}
\newcommand{\bC}{\bmath{C}}
\newcommand{\bE}{\bmath{E}}
\newcommand{\bF}{\bmath{F}}
\newcommand{\bG}{\bmath{G}}
\newcommand{\br}{\bmath{r}}
\newcommand{\bg}{\bmath{g}}
\newcommand{\bd}{\bmath{d}}
\newcommand{\bv}{\bmath{v}}

\newcommand{\bn}{\bmath{n}}
\newcommand{\by}{\bmath{y}}
\newcommand{\bJ}{\bmath{J}}
\newcommand{\bD}{\bmath{D}}
\newcommand{\bH}{\bmath{H}}
\newcommand{\bN}{\bmath{N}}
\newcommand{\bM}{\bmath{M}}
\newcommand{\bO}{\bmath{O}}
\newcommand{\bP}{\bmath{P}}
\newcommand{\bQ}{\bmath{Q}}

\newcommand{\bI}{\bmath{I}}
\newcommand{\ba}{\bmath{a}}
\newcommand{\bb}{\bmath{b}}
\newcommand{\bx}{\bmath{x}}

\newcommand{\bmJ}{\bmath{\mathcal{J}}}
\newcommand{\bmH}{\bmath{\mathcal{H}}}
\newcommand{\bzero}{\bmath{0}}

\newcommand{\conj}[1]{\overline{#1}}

\title[Redundant calibration and Complex Optimization]{Redundant interferometric calibration as a complex optimization problem}
\author[T.L.~Grobler et al.]{T.L.~Grobler$^{1,2}$\thanks{E-mail:tlgrobler@sun.ac.za}, G.~Bernardi$^{2,3,4}$, J.S.~Kenyon$^{2}$, A.R.~Parsons$^{5,6}$ and O.M.~Smirnov$^{2,3}$\\
$^{1}$Dept of Mathematical Sciences, Computer Science Division, Stellenbosch University, Private Bag X1, 7602 Matieland, South Africa\\
$^{2}$Department of Physics and Electronics, Rhodes University, PO Box 94, Grahamstown, 6140, South Africa\\
$^{3}$SKA SA, 3rd Floor, The Park, Park Road, Pinelands, 7405, South Africa\\
$^{4}$INAF-Istituto di Radioastronomia, via Gobetti 101, 40129, Bologna, Italy\\
$^{5}$Dept. of Astronomy, University of California, Berkeley, CA 94720, USA\\
$^{6}$Radio Astronomy Lab., U. California, Berkeley CA 94720, USA}
\begin{document}


\pagerange{\pageref{firstpage}--\pageref{lastpage}} \pubyear{2002}

\maketitle

\label{firstpage}

\begin{abstract}
Observations of the redshifted 21-cm line from the epoch of reionization have recently motivated the construction of low 
frequency radio arrays with highly redundant configurations. These configurations provide an alternative calibration strategy - 
''redundant calibration" - and boosts sensitivity on specific spatial scales. In this paper, we formulate 
calibration of redundant interferometric arrays as a complex optimization problem. 
We solve this optimization problem via the Levenberg-Marquardt algorithm. This calibration approach is more 
robust to initial conditions than current algorithms and, by leveraging an approximate matrix inversion, allows for further optimization and an efficient implementation (``redundant \textsc{StEfCal}"). We also investigated using the preconditioned conjugate gradient method as an alternative to the approximate matrix inverse, but found that its computational performance is not competitive with respect to ``redundant \textsc{StEfCal}". The efficient implementation of this new algorithm is made publicly available.
\end{abstract}

\begin{keywords}
instrumentation: interferometers -- methods: data analysis -- methods: numerical --techniques: interferometric -- cosmology: observations
\end{keywords}

\section{Introduction}
The quest for the redshifted 21-cm line from the Epoch of Reionization (EoR) is a frontier of modern observational cosmology and has motivated the construction of a series of new interferometric arrays operating at low frequencies over the last decade. EoR measurements are challenging because of the intrinsic faintness of the cosmological signal \citep[see, for instance,][for recent reviews]{Furlanetto2016,McQuinn2016} buried underneath foreground emission which is a few orders of magnitude brighter than the EoR anywhere in the sky \citep[e.g.,][]{Bernardi2009,Bernardi2010,Ghosh2012,Dillon2014,Parsons2014}. 
As miscalibrated foreground emission leads to artifacts that can jeopardize the EoR signal \citep[e.g.,][]{grobler2014,barry2016,ewall-wice2016}, an exceptional interferometric calibration is required. Such calibration needs an accurate knowledge of both the sky emission and the instrumental response \citep[e.g.,][]{Smirnov2011c}.
A skymodel is required for standard calibration as it would be an ill-posed problem if one tries to directly solve for all of the unknowns, i.e. the antenna gains and the uncorrupted 
visibilities, without first attempting to simplify the calibration problem. When the uncorrupted visibilities, however, are predicted from a predefined skymodel the problem simplifies and becomes solvable.
In contrast, if an array is deployed in a redundant configuration, i.e. with receiving elements placed on regular grids so that multiple baselines measure the same sky brightness emission,
we circumvent the need for a skymodel altogether. This is true, since redundancy leads to fewer unknowns (i.e. the number of unique 
$uv$-modes) and if the array is redundant enough it reduces the number of unknowns to such an extend that the calibration problem becomes solvable without having to predict the uncorrupted visibilities from a skymodel. Redundancy, therefore, is a promising path to achieve highly accurate interferometric calibration \citep[][]{Noordam1982,Wieringa1992,Pearson1984,Liu2010,Noorishad2012,Marthi2014,Sievers2017}. 
Moreover, redundant configurations provide a sensitivity boost on particular spatial scales that can be tuned to be the highest signal--to--noise ratio (SNR) EoR modes \citep[][]{Parsons2012,Dillon2016}. These reasons motivated the design and deployment of EoR arrays in redundant configurations like the MIT-EoR \citep{Zheng2014}, the Precision Array to Probe the Epoch of Reionization \citep[PAPER,][]{Ali2015} and the Hydrogen Epoch of Reionization Array \citep[HERA,][]{deboer2017}. 

In this paper we, present a new algorithm to calibrate redundant arrays based on the complex optimization formalism recently introduced by \cite{Smirnov2015}. 
With respect to current algorithms, it is more robust to initial conditions, while remaining comparatively fast. We also show that given certain approximations this new algorithm reduces to the redundant calibration equivalent of the \textsc{StEfCal} algorithm \citep{Salvini2014}.
We investigate the speed-up that the preconditioned conjugate gradient (PCG) method provides if it is employed by the new algorithm \citep{Liu2010}.
A comparison between the computational complexity of the optimized new algorithm and redundant \textsc{StEfCal} is also performed.

A summary of the notation that is used within the paper is presented in 
Table~\ref{tab:glossary}. We also discuss the overset notation used within the paper in a bit more detail below (we use $x$ as an operand proxy in this paper):
\begin{enumerate}
 \item $\widetilde{x}$ -- This notation is used to denote a new scalar value which was derived from the scalar $x$ using a proper mathematical definition.  
 \item $\conj{x}$ -- This notation denotes the conjugation of its operand (the conjugation of $x$). For the readers convenience we will redefine this operator when it is
 first used witin the paper.
 \item $\widehat{x}$ -- This notation denotes that the quantity is an estimated value.
 \item $\breve{\mathbf{x}}$ -- This notation denotes an augmented vector, i.e. $\breve{\mathbf{x}} = [\mathbf{x}^T,\conj{\mathbf{x}}^T]^T$. 
\end{enumerate}

The paper is organized as follows: we review the complex calibration formalism by \citet{Smirnov2015} in Section~\ref{sec:sky_wirtinger}, we extend it to the redundant case in Section~\ref{sec:red_wirtinger}, we present some computational complexity results in Section~\ref{sec:pcg} and we present our conclusions in Section~\ref{sec:conclusions}.

\section{Wirtinger calibration}
\label{sec:sky_wirtinger}

\begin{table}
\centering
\caption{Notation and frequently used symbols.}
\begin{tabular}{|l l|} 
\hline
$x$, $\mathbf{x}$ and $\mathbf{X}$ & scalar, vector and matrix\\
$\mathbb{N}$ & set of natural numbers\\
$\mathbb{C}$ & set of complex numbers\\
$\bI$ & Identity matrix\\
$\alpha_{pq}$, $\phi_{pq}$, $\zeta_{pq}$, $\xi_{pq}$ and & indexing functions\\
$\psi_{pq}:\mathbb{N}^2\rightarrow\mathbb{N}$ &\\
$\conj{x}$ & conjugation\\
$\conj{\mathbf{x}},\conj{\mathbf{X}}$ & element-wise conjugation\\
$x^{2}$ & square\\
$\|\mathbf{x}\|_F$ & Frobenius norm\\
$\mathbf{X}^{-1}$ & matrix inversion\\
$\odot$ & Hadamard product\\
&(element-wise product)\\
$\mathbf{X}^H$ & Hermitian transpose\\
$\mathbf{x}^T,\mathbf{X}^T$ & transpose\\
$<\mathbf{x}>_x$ & averaging over $x$\\
$\frac{\partial}{\partial z}$ & Wirtinger derivative\\
$\widetilde{x}$ & a new quantity derived from $x$\\
$\widehat{x}$ & estimated quantity\\
$\breve{\mathbf{x}}$ & augmented vector $[\mathbf{x}^T,\conj{\mathbf{x}}^T]^T$\\
$i$ & $\sqrt{-1}$\\
$N$,$B$,$L$ and $P$ & number of antennas, baselines,\\
&redundant groups and parameters\\
$\bd$, $\bv$, $\bg$, $\by$, $\br$ and $\bn$& data, predicted, antenna,\\
&true visibility, residual\\
&and noise vector.\\ 
$\bz$ & $\bz = [\bg^T,\by^T]^T$\\
$\Lambda$ & objective function\\
$\lambda$ and $\rho$ & damping factor and $\rho = \frac{1}{1+\lambda}$\\
$\bJ$ & Jacobian matrix\\
$\bH$ & $\bJ^H\bJ$ (Hessian matrix)\\
$\bmH$ & modified Hessian matrix\\
$[\mathbf{X}]_{ij}$& element $ij$ of matrix $\mathbf{X}$\\
$x_{ij}$ & composite antenna index\\
$x_i$ & antenna or redundant group index\\
$x^k$ & iteration number\\
$\bM$ & preconditioner matrix\\
$m$ & number of non-zero entries\\
&in a matrix\\
$\kappa$ & spectral condition number\\
&of a matrix\\
$\gamma$ & sparsity factor of a matrix\\
$\Delta x$ & parameter update\\
$\exists!$ & there exists a unique\\
\hline
\end{tabular}
\label{tab:glossary}
\end{table}

\begin{table}
\centering
\caption{The dimensions of the Jacobian matrices, and their respective sub-matrices, defined in Section~\ref{sec:sky_wirtinger} and Section~\ref{sec:red_wirtinger}.}
\begin{tabular}{|c c|} 
\hline
Matrix & Dimension\\
\hline
\hline
\multicolumn{2}{c}{Wirtinger Calibration}\\
\hline
$\bJ$ & $2B \times 2N$ \\
$\bmJ_1$ & $B \times N$\\
$\bmJ_2$ & $B \times N$\\
\hline
\hline
\multicolumn{2}{c}{Redundant Wirtinger Calibration}\\
\hline
$\bJ$ & $2B \times P$ \\
$\bmJ_1$ & $B \times (N+L)$\\
$\bmJ_2$ & $B \times (N+L)$\\
\hline
\end{tabular}
\label{tab:matrix_dimensions_main}
\end{table}

In a radio interferometer, the true sky visibilities $y_{pq}$ measured by a baseline formed by antenna $p$ and $q$ are always ``corrupted" by the non-ideal response of the receiver, which is often incorporated into a single, receiver-based complex number $g$ (i.e. an antenna gain). The observed visibility $d_{pq}$ is therefore given by \citep{ME1,ME2,RRIME1}
\begin{equation}
\label{eq:vis_definition}
d_{pq} = g_{p}\conj{g_q} \, y_{pq} + n_{pq},
\end{equation}
where $\conj{x}$ indicates complex conjugation and $n_{pq}$ is the thermal noise component. 
The real and imaginary components of the thermal noise are normally distributed with a mean of zero and a
standard deviation $\sigma$:
\begin{equation}
\sigma \propto \frac{T_{\textrm{sys}}}{\sqrt{\Delta \nu \tau}},
\end{equation}
where $T_{\textrm{sys}}$ is equal to the system temperature, $\Delta \nu$ is the observational bandwidth and $\tau$ is the integration time per visibility.

Considering the number of visibilities $B$ (i.e. baselines) measured by an array of $N$ elements $B = \frac{N^2-N}{2}$, equation~\ref{eq:vis_definition} can be expressed in the following vector form:
\begin{equation}
\label{eq:vis_linear_definition}
\bd = \bv + \bn, 
\end{equation}
where 
\begin{align}
 \left [ \bd \right]_{\alpha_{pq}} &= d_{pq}, & \left [ \bv \right ]_{\alpha_{pq}} &= v_{pq}=g_p y_{pq} \conj{g_q},\nonumber\\
 \left [ \bn \right ]_{\alpha_{pq}} &= n_{pq}, &  &\label{eq:vec_linear_definitions}
\end{align}
and 
\begin{equation}
\alpha_{pq} =
\begin{cases}
(q-p) + (p-1)\left (N-\frac{1}{2}p \right ) & \textrm{if}~p<q\\
0 & \textrm{otherwise}
\end{cases}.
\end{equation}
The function $\alpha_{pq}$ therefore maps composite antenna indexes to unique single indexes, i.e:
\begin{equation}
\{\alpha_{12},\alpha_{13},\cdots,\alpha_{N-1N}\} = \{1,2,\cdots,B\} 
\end{equation}
The vectors in equation~\ref{eq:vec_linear_definitions} are column vectors of size $B$ (i.e. $p<q$). It is important to point out here that 
all of the mathematical definitions in this paper assumes that we are only considering composite indices belonging to the set $\{rs|r<s\}$. For the sake of mathematical 
completeness, however, $\alpha_{pq}$ is defined for composite indices which do not belong to this set. Also note that, while the indexing function $\alpha_{pq}$ does not attain 
the value zero in practice the zero indexing value does play a very important role in some of the definitions used within this paper (see for example Eq.~\ref{eq:D_mat}).

Radio interferometric calibration aims to determine the best estimate of $\bg = [g_1,g_2,\cdots,g_N]^T$ in order to correct the data and, following equation~\ref{eq:vis_linear_definition}, can be formulated as a non linear least-squares optimization problem:
\begin{equation}
\label{eq:least_squares}
\min_{\bg} \Lambda(\bg) = \min_{\bg} \|\br\|_F^2 = \min_{\bg} \|\bd - \bv(\bg)\|_F^2, 
\end{equation}
where $\Lambda$ is the objective function, $\br$ is the residual vector and $\|\mathbf{x}\|_F$ denotes the Frobenius norm. In standard interferometric calibration, $y_{pq}$ is assumed to be known at some level, for instance through the observation of previously known calibration sources. 

Non-linear least-squares problems are generally solved by using gradient-based minimization algorithms (i.e. Gauss--Newton (GN) -- or Levenberg--Marquardt (LM)) that require the model ($\bv$ in equation~\ref{eq:least_squares}) to be differentiable towards each parameter. 
When the least squares problem is complex, it becomes less straightforward to apply these gradient-based minimization methods, as
many complex functions are not differentiable if the classic notion of differentiation is used, i.e. $\frac{\partial \conj{z}}{\partial z}$ does not exist if $z \in \mathbb{C}$.

In order to circumvent the differentiability conundrum associated with complex least squares problems, standard interferometric calibration divides the complex optimization problem into its real and imaginary parts and solves for the real and imaginary parts of the unknown model parameters separately. \citet{Smirnov2015} showed, however, that this approach is not needed if complex calculus \citep{Wirtinger1927} is adopted. The Wirtinger derivatives are defined as:
\begin{align}
\label{eq:wir}
\frac{\partial}{\partial z} &= \frac{1}{2}\left ( \frac{\partial}{\partial x} -  i \frac{\partial}{\partial y} \right ),&\frac{\partial}{\partial \conj{z}} &= \frac{1}{2}\left ( \frac{\partial}{\partial x} +  i \frac{\partial}{\partial y} \right ), 
\end{align}
which lead to the following relations:
\begin{align}
\label{eq:wir_z}
\frac{\partial z}{\partial z} & = 1, & \frac{\partial \conj{z}}{\partial z}&=0, & \frac{\partial z}{\partial \conj{z}} & = 0, & \frac{\partial \conj{z}}{\partial \conj{z}}&=1.
\end{align}
If the gradient operator is defined using equation~\ref{eq:wir}, the model $\bv$ now becomes analytic in both $\bg$ and $\conj{\bg}$ and equation \ref{eq:wir_z} can be used to derive the complex variants of the real-valued GN and LM algorithms. In the complex GN and LM algorithms, complex parameters and their conjugates are treated as separate variables.

Assuming that $y_{pq}$ is known, equation~\ref{eq:least_squares} is recast as \citep{Smirnov2015}:
\begin{equation}
\label{eq:least_squares_augmented}
\min_{\breve{\bg}} \Lambda(\breve{\bg}) = \min_{\breve{\bg}} \|\breve{\br}\|_F^2 = \min_{\breve{\bg}} \|\breve{\bd} - \breve{\bv}(\breve{\bg})\|_F^2, 
\end{equation} 
where $\breve{\br} = [\br^T,\conj{\br}^T]^T$, $\breve{\bd} = [\bd^T,\conj{\bd}^T]^T$, $\breve{\bv} = [\bv^T,\conj{\bv}^T]^T$ and $\breve{\bg} = [\bg^T,\conj{\bg}^T]^T$.

The complex GN update is therefore defined as:
\begin{equation}
\label{eq:GN_update_skymodel}
 \Delta \breve{\bg} = (\bJ^H\bJ)^{-1}\bJ^H\breve{\br},
\end{equation}
with 
\begin{equation}
\label{eq:Jacobian_skymodel}
\bJ = \begin{bmatrix}
       \bmJ_1 & \bmJ_2\\
       \conj{\bmJ}_2 & \conj{\bmJ}_1 
      \end{bmatrix},
\end{equation}
and
\begin{align}
\label{eq:jac_entries}
[\bmJ_1]_{\alpha_{pq},i} &= \frac{\partial v_{pq}}{\partial g_i}, & [\bmJ_2]_{\alpha_{pq},i} &= \frac{\partial v_{pq}}{\partial \conj{g}_i}. 
\end{align}
The matrix $\bJ$ is generally referred to as the Jacobian\footnote{a matrix composed of first-order partial derivatives} matrix. 

The complex LM update is very similar, the major difference being the introduction of a damping parameter, $\lambda$:
\begin{equation}
\label{eq:LM_update_skymodel}
\Delta \breve{\bg} = (\bJ^H\bJ + \lambda\bD)^{-1}\bJ^H\breve{\br},
\end{equation}
where $\bD=\bI\odot\bJ^H\bJ$. 

In this paper, we use single subscript indices (e.g. $i$) to refer to a specific antenna (or as will become apparent later a specific redundant group) and composite subscript indices (e.g. $pq$) to refer to 
a specific baseline. If these indices appear in the definition of matrix elements then their allowed ranges are determined by the dimension of the matrix that 
is being defined (see Table~\ref{tab:matrix_dimensions_main}). Furthermore, the identity matrix is denoted by $\bI$. Moreover, the Hadamard\footnote{element-wise product} product and the Hermition transpose are denoted by $\odot$ and $\mathbf{X}^H$ respectively \citep{Liu2008}. Note the use of the Wirtinger derivatives in equation~\ref{eq:jac_entries}.
We will refer to $\bJ^H\bJ$ as the Hessian\footnote{a square matrix composed of second-order partial derivatives} matrix $\bH$ and to $\bJ^H\bJ + \lambda\bD$ as the modified Hessian matrix $\bmH$ throughout this paper \citep{Madsen:LM}. 

Equation~\ref{eq:GN_update_skymodel} or~\ref{eq:LM_update_skymodel} can now be used iteratively to update the parameter vector $\breve{\bg}$:
\begin{equation}
\label{eq:update_skymodel}
\breve{\bg}_{k+1} = \breve{\bg}_{k} + \Delta \breve{\bg}_{k},
\end{equation}
until convergence is reached.

In the case of the GN algorithm, the parameter update step simplifies and becomes \citep{Smirnov2015}
\begin{equation}
\label{eq:one_half}
\breve{\bg}_{k+1} = (\bJ^H\bJ)^{-1}\bJ^H\breve{\bd} + \frac{1}{2}\breve{\bg}_{k}. 
\end{equation}

\citet{Smirnov2015} realized that the diagonal entries of the Hessian matrix $\bH$ are much more significant than its off-diagonal entries, i.e. $\bH$ is nearly diagonal. By approximating $\bH$ by its diagonal and substituting the approximate Hessian matrix the LM parameter update step becomes \citep{Smirnov2015}:
\begin{align}
\breve{\bg}_{k+1} &\approx \frac{1}{1+\lambda}\widetilde{\bH}^{-1}\bJ^H\breve{\bd} + \frac{\lambda}{1+\lambda} \breve{\bg}_k, \nonumber \\
 &= \rho \widetilde{\bH}^{-1}\bJ^H\breve{\bd} + (1-\rho)\breve{\bg}_k, 
\label{eq:stef_alpha}  
\end{align}
where $\rho = \frac{1}{1+\lambda}$. Note that equation~\ref{eq:one_half} and equation~\ref{eq:stef_alpha} are   
not dependent on $\breve{\br}$. 

Interestingly enough, if $\lambda = 0$ we obtain the odd parameter update step of \textsc{StEfCal}\footnote{$k\in\{0,2,\cdots\}$}, and if $\lambda=1$ (which corresponds
to $\rho=\frac{1}{2}$) we obtain the even parameter update step of \textsc{StEfCal}\footnote{$k\in\{1,3,\cdots\}$} \citep[\textsc{StEfCal},][]{Mitchell:MWA-cal,Salvini2014}. In the \textsc{StEfCal} algorithm, the measurement equation (equation~\ref{eq:vis_definition}) is linearized by assuming that the gains are known, but that their conjugates are not. Under this assumption, the system of equations become linear and the conjugates of the gains can be obtained in a straightforward manner. Starting from the latest value of the gain conjugates, an updated estimate of the gains can be obtained iteratively until convergence is reached.
Alternating between solving and fixing different sets of parameters (which is exactly what \textsc{StEfCal} does) is referred to 
as the alternating direction implicit (ADI) method. The \textsc{StEfCal} algorithm reduces the computational complexity of calibration from $O(N^3)$ to $O(N^2)$.

\section{Redundant Wirtinger Calibration}
\label{sec:red_wirtinger}
Interferometric baselines are redundant when they sample the exact same visibilities in the $uv$-plane, i.e. if baseline $pq$ and $rs$ are redundant then $y_{pq} = y_{rs}$. 
A redundant array layout allows us to solve for the unknown observed visibilities themselves in
addition to the antenna gains (see equation \ref{eq:vec_linear_definitions}). This is true, since in the case of a redundant layout equation \ref{eq:vec_linear_definitions} is already an overdetermined 
system even before having predicted visibilities from a pre-existing skymodel. 


It is convenient to group redundant visibilities together and label each group using a single index rather than using their antenna pairs as in equation~\ref{eq:vis_definition}. We introduce a function $\phi$ that maps the antenna pair associated with a specific baseline to its corresponding redundant baseline group, i.e. if baseline $pq$ and $rs$ are redundant then $\phi_{pq} = \phi_{rs}$ (implying they belong to the same group). 
To be exact, $\phi$ maps the composite index $pq$ to its group index only if $pq\in\{rs|r\leq s\}$. If $pq\notin\{rs|r\leq s\}$ then the 
composite index $pq$ is mapped to zero. The function $\phi_{pq}$ is, therefore, not symmetric. Equation~\ref{eq:vis_definition} can be re-written for a redundant array as:
\begin{equation}
\label{eq:vis_red}
d_{pq} = g_{p}\conj{g_q}y_{\phi_{pq}} + n_{pq},
\end{equation}
with the same vector form as equation~\ref{eq:vis_linear_definition} if
\begin{align}
 \left [ \bd \right]_{\alpha_{pq}} &= d_{pq}, & \left [ \bv \right ]_{\alpha_{pq}} &= v_{pq}=g_p y_{\phi_{pq}} \conj{g_q},\nonumber\\
 \left [ \bn \right ]_{\alpha_{pq}} &= n_{pq}, &  &\label{eq:vec_definitions}
\end{align}
where the vectors in Equation~\ref{eq:vec_definitions} are column vectors of size $B$ (i.e. $p<q$).

\begin{figure*}
\centering
\subfigure[Hexagonal layout]
{\includegraphics[width=0.32\textwidth]{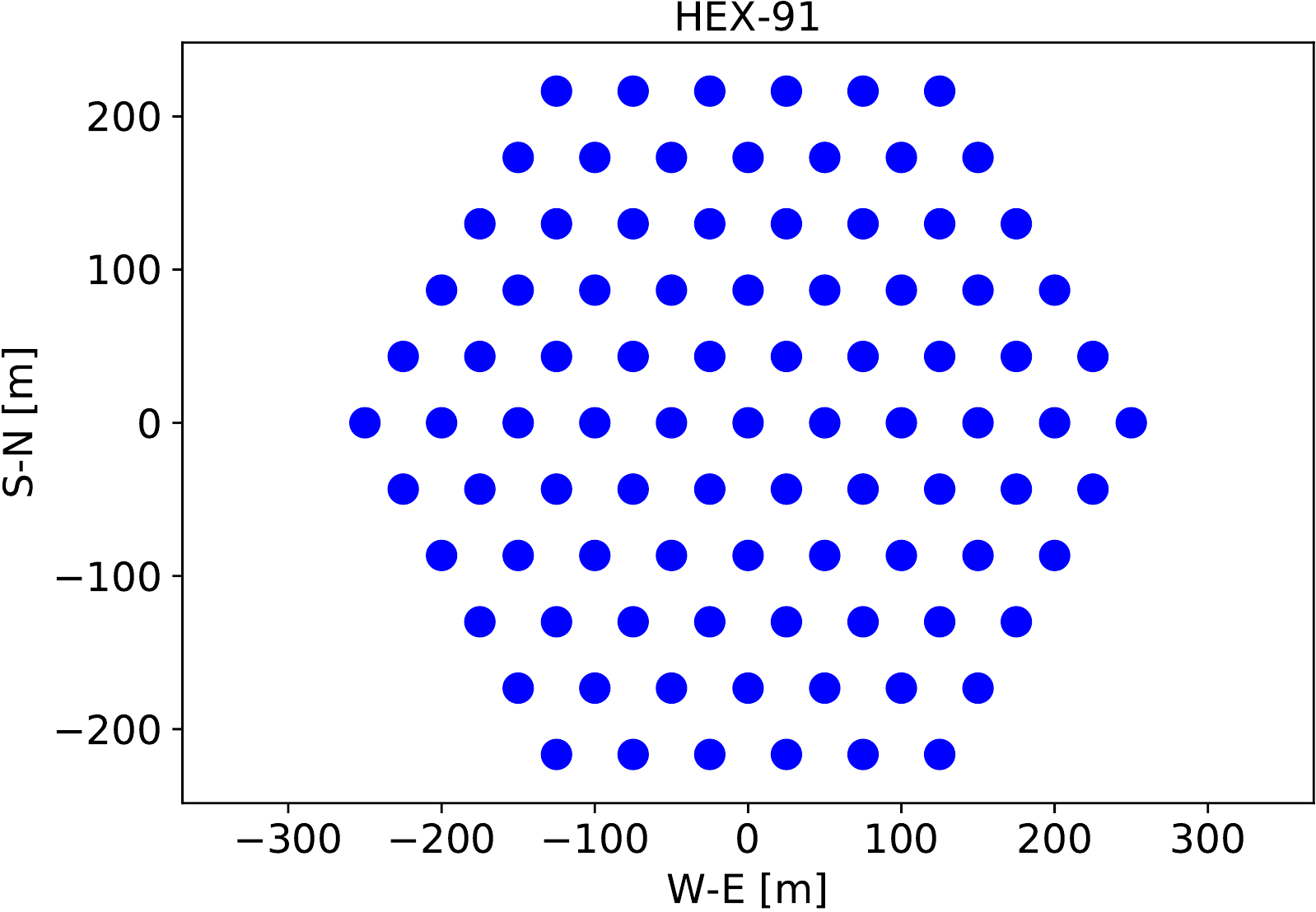}\label{fig:HEX_lay}}
\subfigure[Square layout]
{\includegraphics[width=0.32\textwidth]{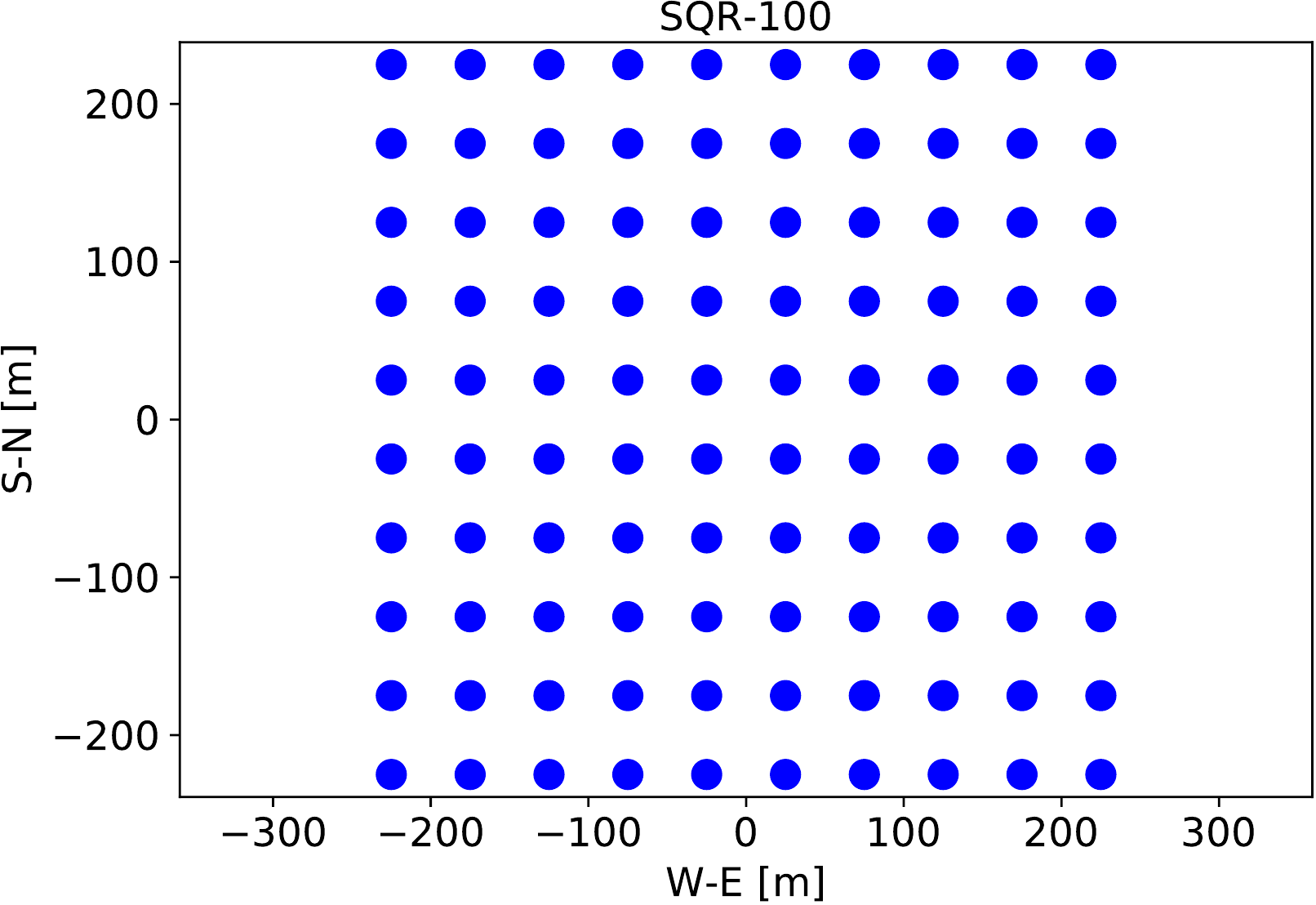}\label{fig:SQR_lay}}
\subfigure[Regular east-west layout]
{\includegraphics[width=0.32\textwidth]{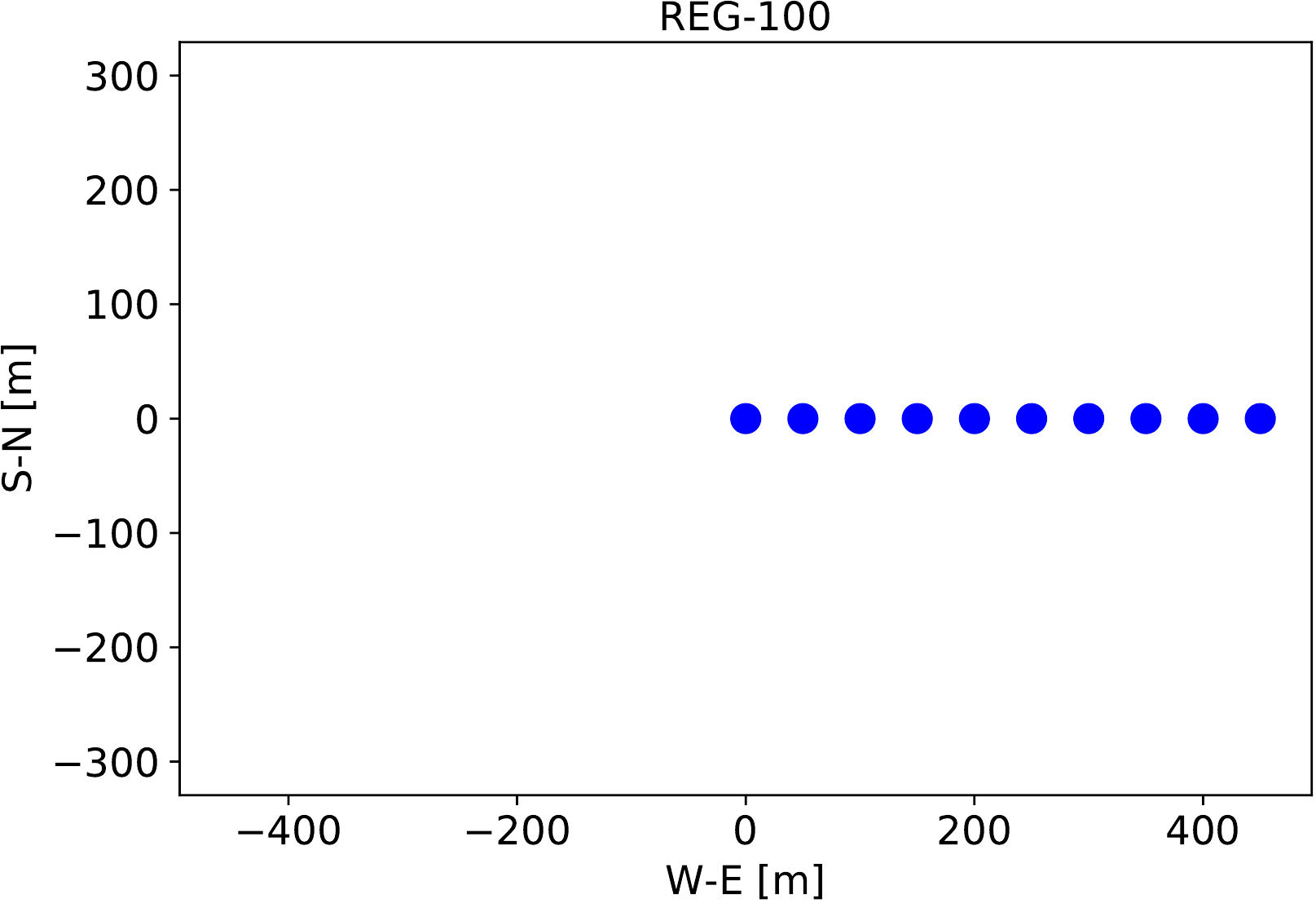}\label{fig:REG_lay}}

\subfigure[Hexagonal: $\zeta_{pq}$]
{\includegraphics[width=0.33\textwidth]{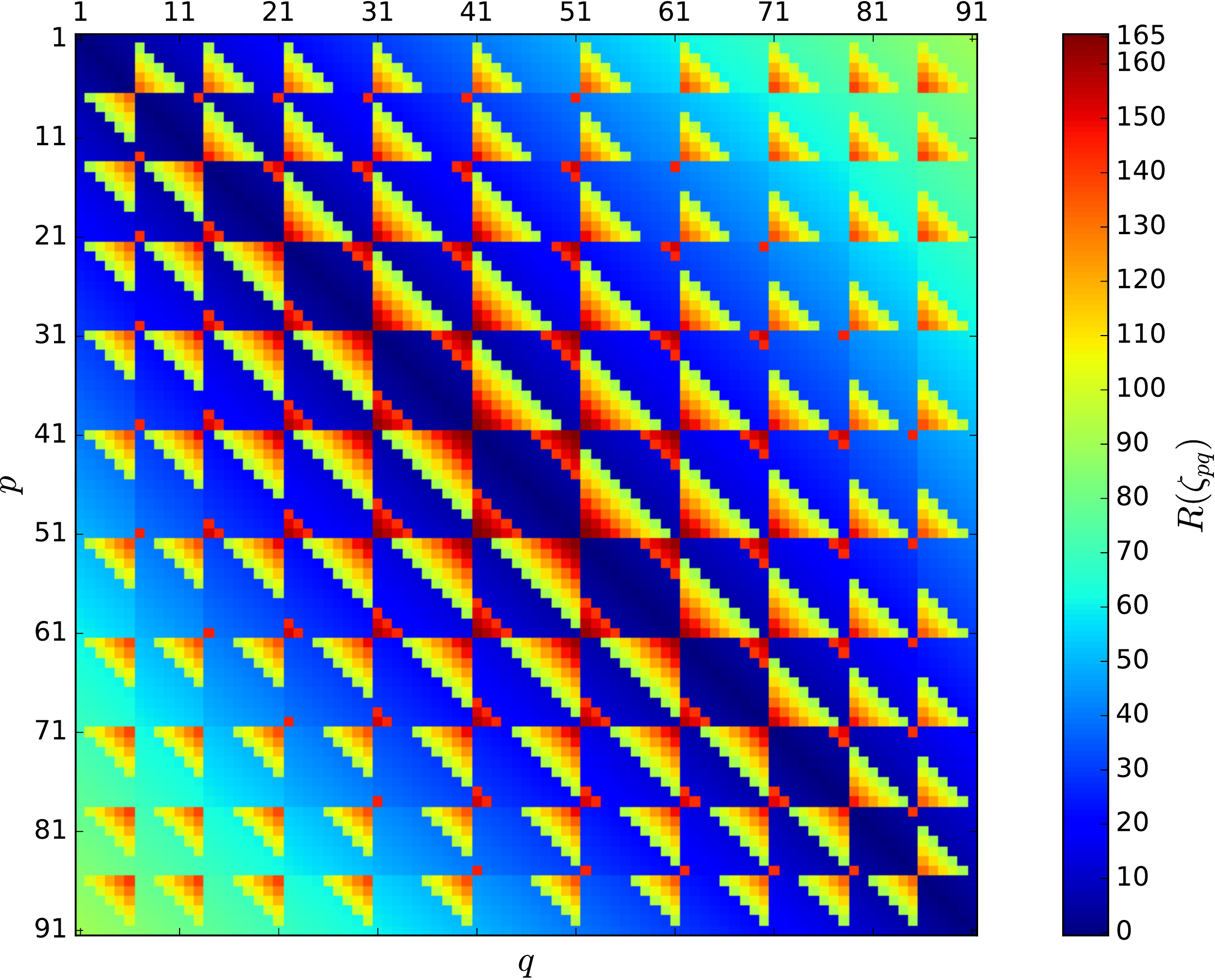}\label{fig:HEX_phi}}
\subfigure[Square: $\zeta_{pq}$]
{\includegraphics[width=0.33\textwidth]{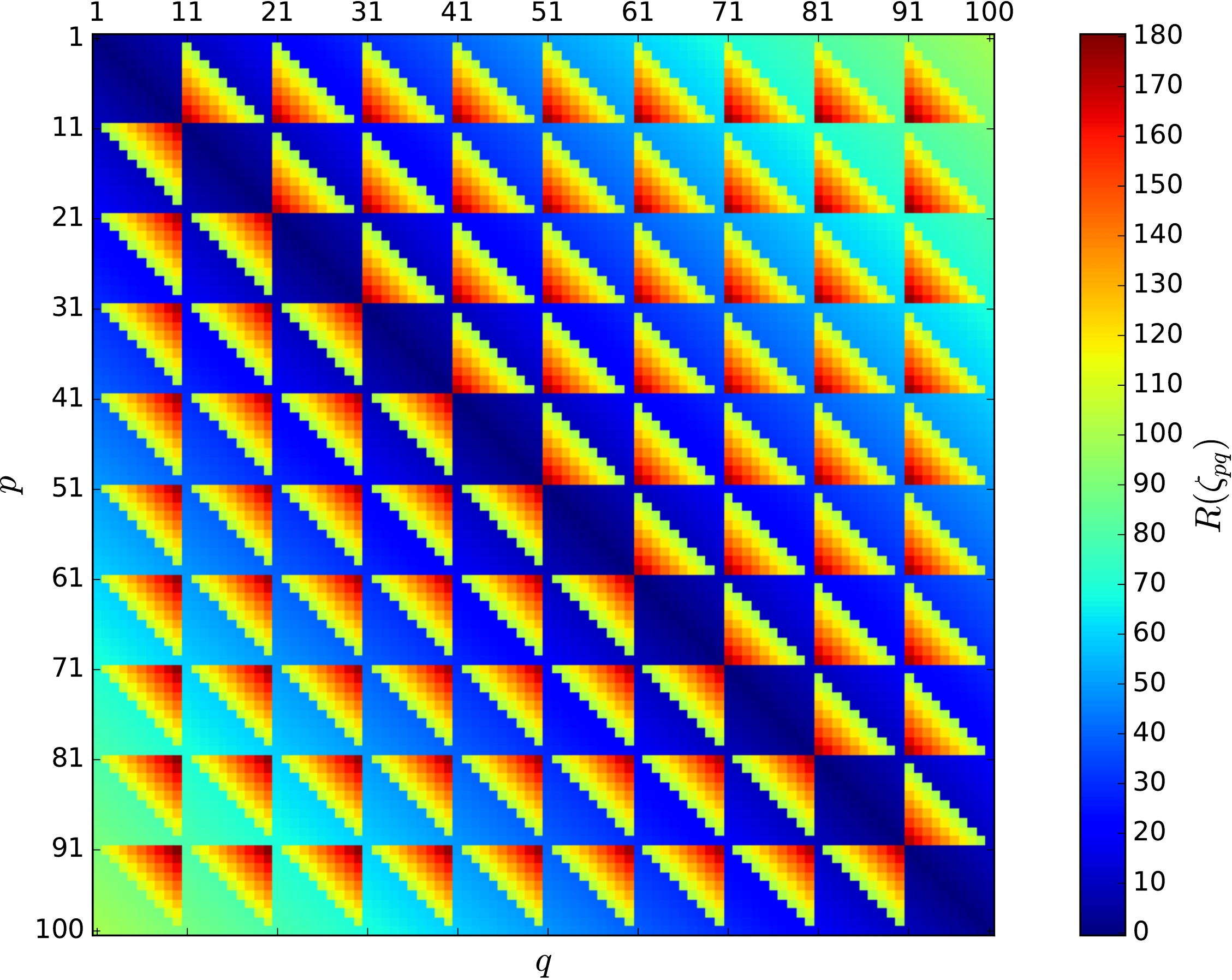}\label{fig:SQR_phi}}
\subfigure[Regular east-west: $\zeta_{pq}$]
{\includegraphics[width=0.33\textwidth]{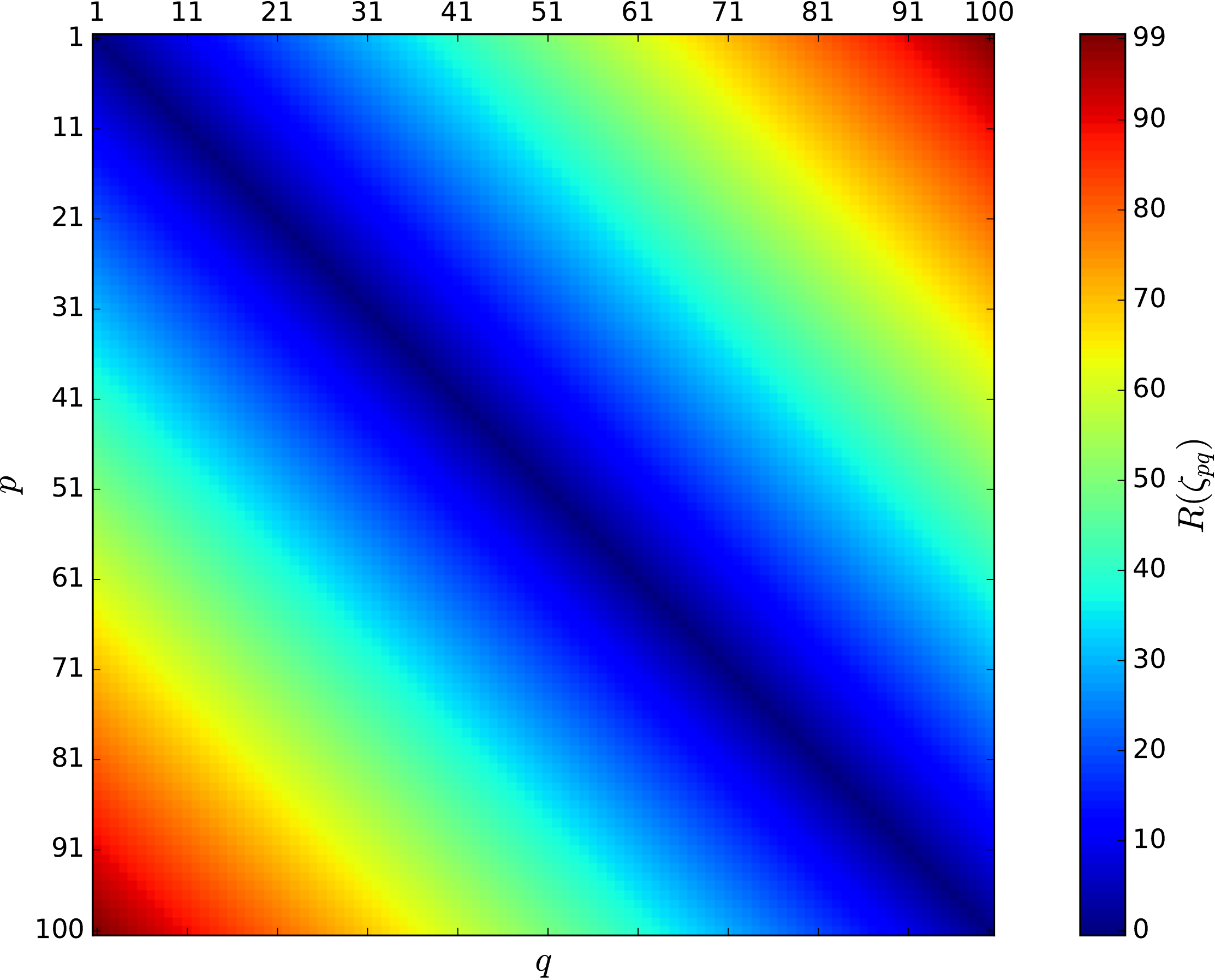}\label{fig:REG_phi}}
\caption{Three different redundant antenna layouts: hexagonal (top left), square (top middle) and regular east-west (top right) with their associated symmetric redundancy geometry functions $\zeta_{pq}$ (bottom panels). We used 91 antennas to construct the hexagonal layout, while 100 antennas were used in the square and east-west layouts. In the case of the east-west layout we only plot the positions of the first ten antennas. The maximal amount of redundant baseline groups $L$ that can be formed for 
the hexagonal, square and east-west layouts are 165, 180 and 99 respectively. The analytic expressions of $L$ for an hexagonal, square and east-west layout are $L = 2N-\frac{1}{2}\sqrt{12N-3}-\frac{1}{2}$,
$L=2N-2\sqrt{N}$ and $L=N-1$ respectively.\label{fig:geometry_function}}
\end{figure*}

We also introduce the following symmetric variant of $\phi_{pq}$:
\begin{equation}
\zeta_{pq} = 
\begin{cases}
\phi_{pq}~\textrm{if}~p \leq q\\
\phi_{qp}~\textrm{if}~p>q
\end{cases},
\end{equation} 
and we will refer to $\zeta_{pq}$ as the symmetric geometric function.
It is possible to construct a simple analytic expression for $\zeta_{pq}$ for an east-west regular array, i.e. $\zeta_{pq} = |q-p|$. It becomes, however, increasingly difficult to construct analytic expressions of $\zeta_{pq}$ for more complicated array layouts. The empirically constructed symmetric geometry functions of three different redundant layouts are displayed in Figure~\ref{fig:geometry_function}. We denote the range of $\zeta_{pq}$ with $\mathcal{R}(\zeta_{pq})$. The maximal element that $\zeta_{pq}$ can ascertain is denoted by $L$ and can be interpreted as the greatest number of unique redundant baseline groups which can be formed for a given array layout. 

We can now formulate redundant calibration as a least-squares problem:
\begin{equation}
\label{eq:least_squares_red}
\min_{\bz} \Lambda(\bz) = \min_{\bz} \|\br\|_F^2 = \min_{\bz} \|\bd - \bv(\bz)\|_F^2, 
\end{equation}
where
\begin{align}
 \bg &=[g_1,\cdots,g_N]^T, & \by &= [y_1,\cdots,y_L]^T,\nonumber\\
 \bz &= [\bg^T,\by^T]^T. &  &\label{eq:parm_definitions}
 \end{align}
The number of model parameters to be solved for is now $P = 2(N+L)$, since redundant calibration is a complex problem. Note that equation~\ref{eq:least_squares_red} is only solvable 
(i.e. the array is redundant enough) if $L+N \leq B$.

In literature, equation~\ref{eq:least_squares_red} is solved by splitting the problem into its real and imaginary parts. The real and imaginary parts of the unknown parameters are then solved for separately \citep{Wieringa1992,Liu2010,Zheng2014}. 
Currently, the above is achieved by using the real-valued GN algorithm \citep{Kurien2016}. 
We, instead, intend to formulate the redundant calibration problem using Wirtinger calculus and recast equation~\ref{eq:least_squares_red} as
\begin{equation}
\label{eq:least_squares_complex}
\min_{\breve{\bz}} \|\breve{\br}\| = \min_{\breve{\bz}} \|\breve{\bd} - \breve{\bv}(\breve{\bz})\|, 
\end{equation}
where $\breve{\bz} = [\bz^T,\conj{\bz}^T]^T$. 

\begin{figure}
\includegraphics[width=0.47\textwidth]{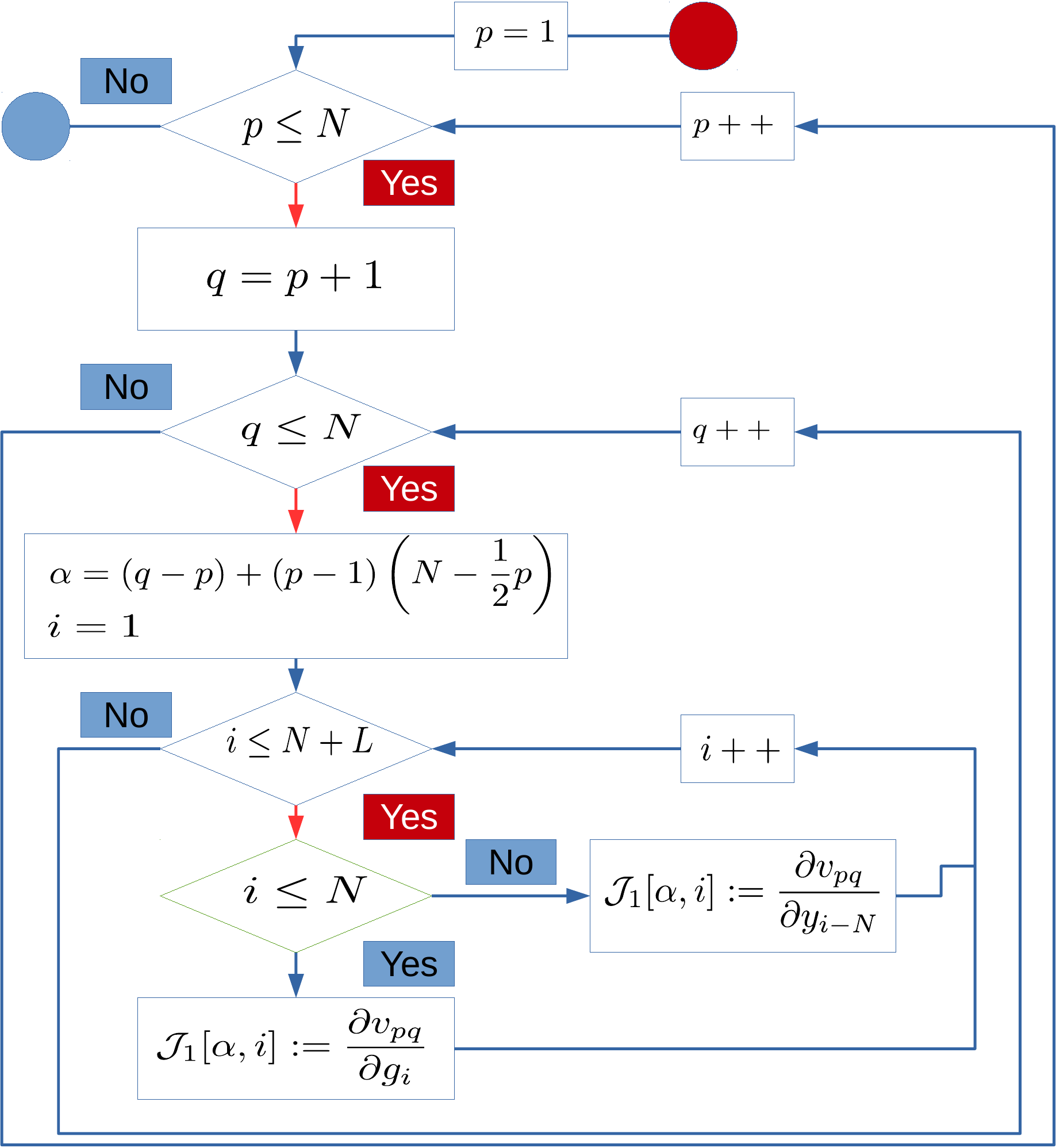}
\caption{A flow chart representing the procedure one would follow to partially construct the Jacobian matrix (i.e. $\bmJ_1$) associated with Wirtinger redundant 
calibration. The red circle represents the start of the flow diagram. The blue circle represents the end of the diagram. Blue diamonds denote loop 
conditionals, while green diamonds denote simple conditionals. The diagram elements following the red arrows just below a loop conditional element all form part 
of the main body of the loop which has the conditional statement of the aforementioned loop conditional element in its definition. The pseudocode associated with this flow diagram is given in Algorithm 1. \label{fig:flow_chart}}
\end{figure}

We derive the complex Jacobian associated with equation~\ref{eq:least_squares_complex} to be:
\begin{equation}
\label{eq:Jacobian}
\bJ = \begin{bmatrix}
       \bmJ_1 & \bmJ_2\\
       \conj{\bmJ}_2 & \conj{\bmJ}_1 
      \end{bmatrix},
\end{equation}
where 
\begin{equation}
\label{eq:j1}
[\bmJ_1]_{\alpha_{pq},i} = \begin{cases} 
     \frac{\partial v_{pq}}{\partial g_i} & \textrm{if}~i\leq N \\
     \frac{\partial v_{pq}}{\partial y_{i-N}} & \textrm{otherwise}  
\end{cases}, 
\end{equation}
and
\begin{equation}
\label{eq:j2}
[\bmJ_2]_{\alpha_{pq},i} = \begin{cases} 
     \frac{\partial v_{pq}}{\partial \conj{g}_i} & \textrm{if}~i\leq N \\
     \frac{\partial v_{pq}}{\partial \conj{y}_{i-N}} & \textrm{otherwise}  
\end{cases}. 
\end{equation}

\begin{algorithm}
\caption{Constructing $\bmJ_1$}\label{euclid}
\begin{algorithmic}[1]
\State $p \gets 1$
\While {$p\leq N$}
\State $q \gets p + 1$
\While {$q\leq N$}
\State $\alpha  \gets (q-p) + (p-1)\left (N-\frac{1}{2} \right )$
\State $i \gets 1$
\While {$i\leq (N + L)$}
\If {$i \leq N$}
\State $\bmJ_1[\alpha,i] \gets \frac{\partial v_{pq}}{\partial g_i}$
\Else
\State $\bmJ_1[\alpha,i] \gets \frac{\partial v_{pq}}{\partial y_{i-N}}$
\EndIf
\State $i\gets i + 1$
\EndWhile
\State $q \gets q + 1$
\EndWhile
\State $p \gets p + 1$
\EndWhile
\end{algorithmic}
\end{algorithm}

Note that we employ the same subscript indexing notation that we used in Section~\ref{sec:sky_wirtinger} in equation~\ref{eq:j1} and equation~\ref{eq:j2}.
To further aid the reader in understanding this indexing notation we refer him/her to Figure~\ref{fig:flow_chart} (also see Algorithm 1) which depicts a flow diagram of the matrix construction procedure 
with which $\bmJ_1$ can be constructed. The range of values the indexes in equation~\ref{eq:j1} and equation~\ref{eq:j2} can attain should also 
be clear to the reader after having inspected Figure~\ref{fig:flow_chart} and Table~\ref{tab:matrix_dimensions_main}. 

We can now calculate the GN and LM updates to be 
\begin{equation}
\label{eq:GN_update}
\Delta \breve{\bz} = (\bJ^H\bJ)^{-1}\bJ^H\breve{\br}
\end{equation}
and 
\begin{equation}
\label{eq:LM_update}
\Delta \breve{\bz} = (\bJ^H\bJ + \lambda\bD)^{-1}\bJ^H\breve{\br},
\end{equation}
respectively. As in Section~\ref{sec:sky_wirtinger}, equation~\ref{eq:GN_update} can be used to iteratively update our parameter vector:
\begin{equation}
\label{eq:update}
\breve{\bz}_{k+1} = \breve{\bz}_{k} + \Delta \breve{\bz}_{k}. 
\end{equation}
The analytic expressions for $\bJ$, $\bH$ and $\bJ^H\breve{\br}$ can be found in Appendix~\ref{sec:analytic}.
Appendix~\ref{sec:analytic} also contains two useful identities involving $\bJ$ and $\bH$. 

In the case of the GN algorithm we can simplify equation~\ref{eq:update} even further. Replacing $\breve{\br}$ with $\breve{\bd}-\breve{\bv}$ in equation~\ref{eq:GN_update} results in
\begin{equation}
\label{eq:temp_GN_eq}
\Delta \breve{\bz} = (\bJ^H\bJ)^{-1}\bJ^H(\breve{\bd}-\breve{\bv}), 
\end{equation}
If we substitute the first identity of equation~\ref{eq:identities} into equation~\ref{eq:temp_GN_eq} and we simplify the 
result we obtain 
\begin{equation}
\label{eq:one_thrid}
\Delta \breve{\bz} = (\bJ^H\bJ)^{-1}\bJ^H\breve{\bd}-\frac{1}{3}\breve{\bz}.
\end{equation}
If we use the above simplified update in equation~\ref{eq:update} it reduces to
\begin{equation}
\label{eq:two_thrids}
\breve{\bz}_{k+1} = (\bJ^H\bJ)^{-1}\bJ^H\breve{\bd} + \frac{2}{3}\breve{\bz}_{k}. 
\end{equation}
Equation~\ref{eq:two_thrids} is the redundant equivalent of equation~\ref{eq:one_half} and it shows us that in the case of redundant calibration we can calculate the GN parameter update step without calculating the residual.  

\begin{figure*}
\centering
\subfigure[Regular layout]{\includegraphics[width=0.47\textwidth]{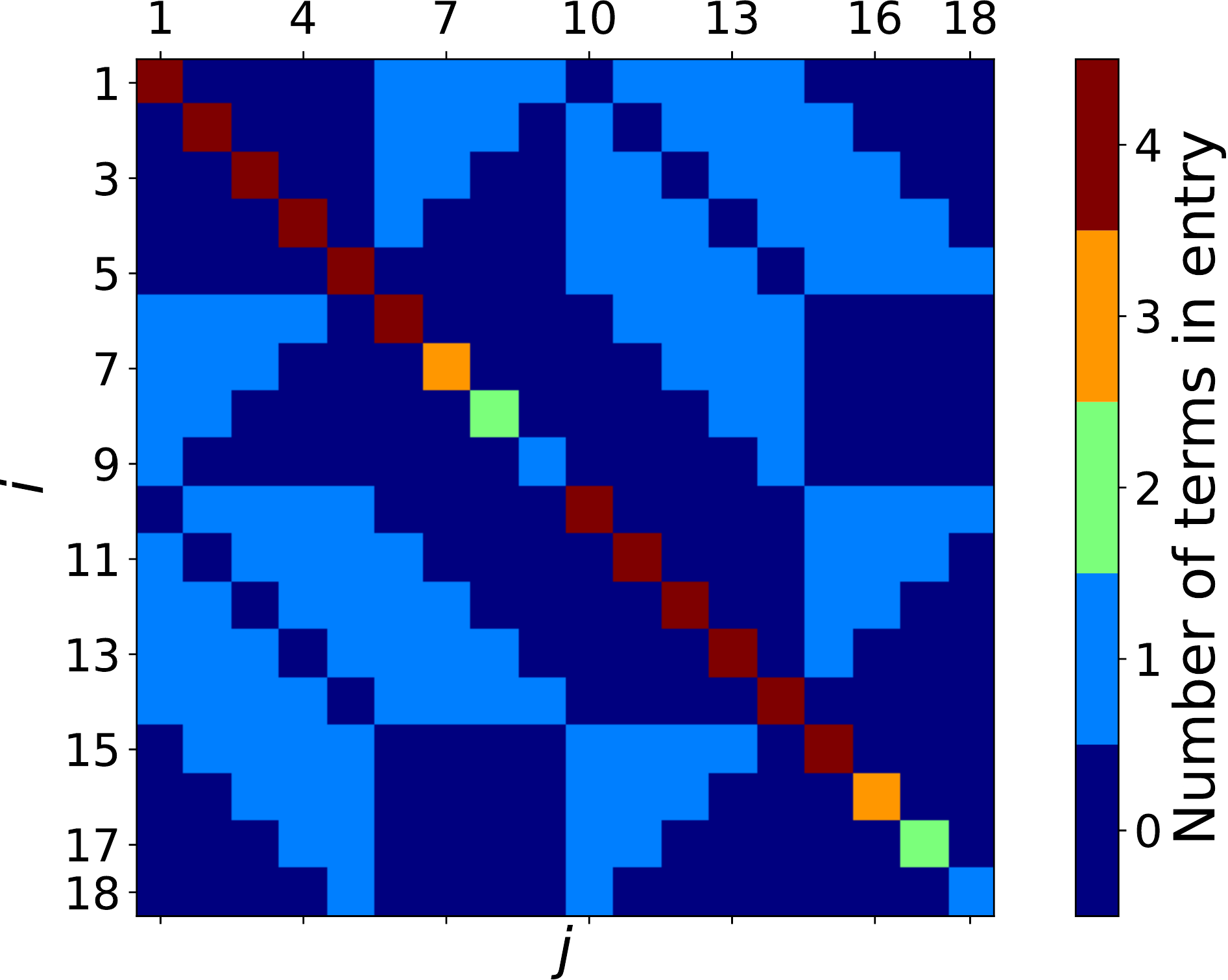}\label{fig:hessian_reg}}
\subfigure[Hexagonal layout]{\includegraphics[width=0.47\textwidth]{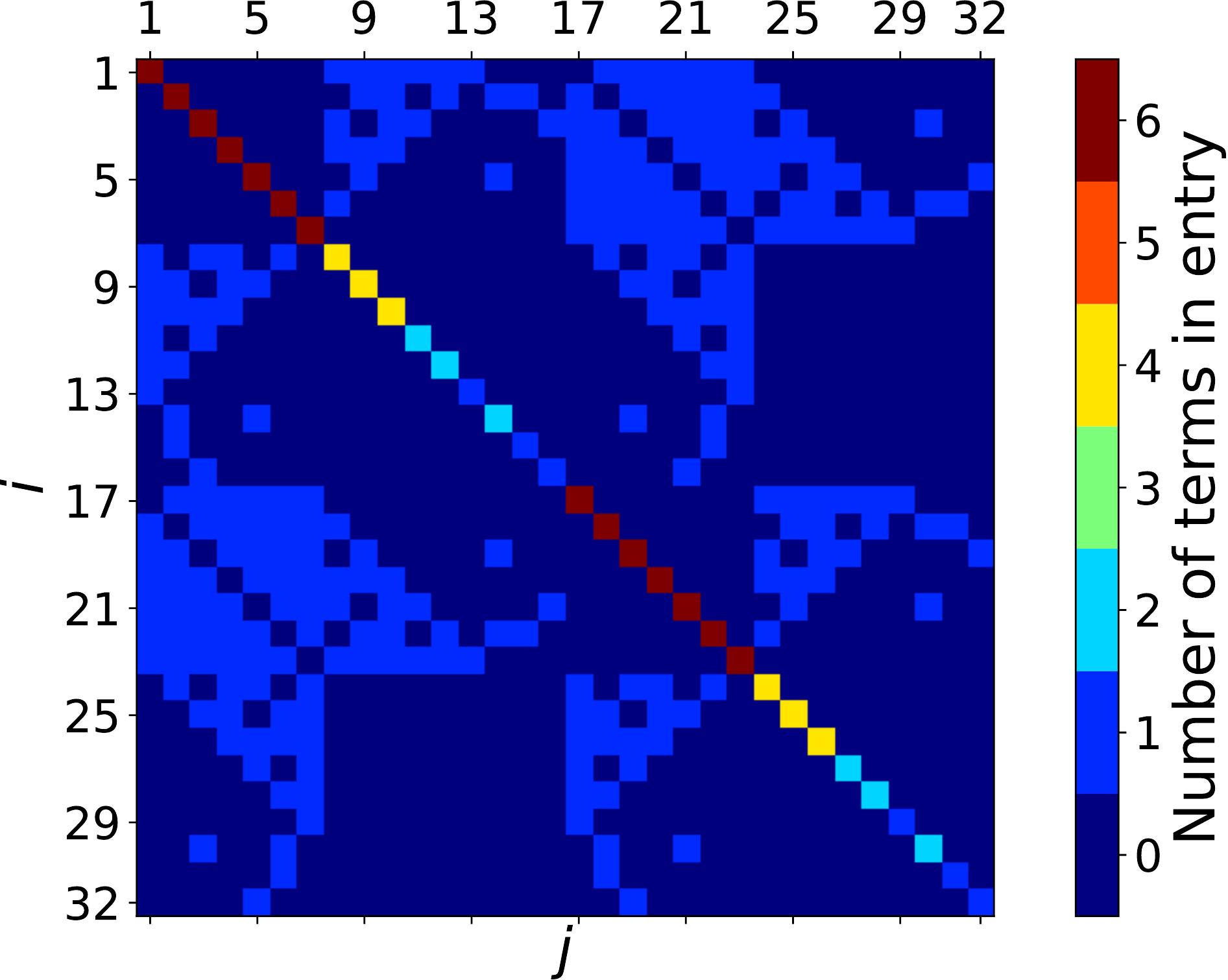}\label{fig:hessian_hex}}
\caption{The number of analytic terms out of which the entries of the Hessian $\bH$ consist for two different geometric layouts, namely a regular east-west grid with $N = 5$ (left panel) and 
a hexagonal grid with $N = 7$. The diagonal entries of these two Hessians are clearly more significant than their off-diagonal entries. Moreover, these two Hessians also 
contain many zero-entries. Note that the locations of the zero-entries are dependent on the geometry of the array layout.
\label{fig:hessian}} 
\end{figure*}

Figure~\ref{fig:hessian} shows that the Hessian matrix $\bH$ is nearly diagonal and sparse for both the regular east-west and hexagonal layouts we considered. We therefore follow the approach of \citet{Smirnov2015} and approximate the Hessian matrix $\bH$ with its diagonal. If we substitute $\bJ^H\bJ$ with $\widetilde{\bH}=\bH\odot\bI$ and replace $\breve{\br}$ with $\breve{\bd} - \breve{\bv}$ in equation~\ref{eq:GN_update} we obtain:
\begin{equation}
\label{eq:appr_GN}
 \Delta \breve{\bz} \approx \widetilde{\bH}^{-1}\bJ^H(\breve{\bd}-\breve{\bv})
\end{equation}
Utilizing the second identity in equation~\ref{eq:identities} allows us to simplify equation~\ref{eq:appr_GN} to
\begin{equation}
  \Delta \breve{\bz} \approx \widetilde{\bH}^{-1}\bJ^H\breve{\bd}-\breve{\bz},
\end{equation}
which leads to
\begin{equation}
 \breve{\bz}_{k+1} \approx \widetilde{\bH}^{-1}\bJ^H\breve{\bd}.
\end{equation}
Using equation~\ref{eq:LM_update}, we follow the same procedure and obtain a similar result for LM
\begin{align}
\breve{\bz}_{k+1} &\approx \frac{1}{1+\lambda}\widetilde{\bH}^{-1}\bJ^H\breve{\bd} + \frac{\lambda}{1+\lambda} \breve{\bz}_k,\label{eq:lambda}\\
 &= \rho \widetilde{\bH}^{-1}\bJ^H\breve{\bd} + (1-\rho)\breve{\bz}_k. \label{eq:alpha}  
\end{align}
The analytic expression of $\bJ^H\breve{\bd}$ will be very similar to the analytic 
expression of $\bJ^H\breve{\br}$, the only difference being that in equation~\ref{eq:ab} the letter $r$ would be replaced by a $d$. If we substitute the analytic expression
of $\bJ^H\breve{\bd}$ and $\widetilde{\bH}^{-1}$ (which can easily be constructed using Appendix~\ref{sec:analytic}) into equation~\ref{eq:alpha} we obtain the following two update rules:
\begin{equation}
\label{eq:g_update}
g_{i}^{k+1} = \rho \frac{\sum_{j\neq i} g_j^k \widetilde{y}_{ij}^{~\!\!k} d_{ij}}{\sum_{j\neq i} |g_j^k|^2|y_{\zeta_{ij}}^k|^2} + (1-\rho) g_i^k, 
\end{equation}
and
\begin{equation}
\label{eq:y_update}
y_{i}^{k+1} = \rho \frac{\sum_{rs \in \mathcal{RS}_i} \conj{g}_r^k g_s^k d_{rs}}{\sum_{rs \in \mathcal{RS}_i}|g_r^k|^2|g_s^k|^2} + (1-\rho) y_i^k. 
\end{equation}
The the index set $\mathcal{RS}_i$ and the quantity $\widetilde{y}_{ij}$ are defined in equation~\ref{eq:RS} and equation~\ref{eq:y_tilde}, respectively. 
The computational complexity of inverting $\widetilde{\bH}$ is $O(P)$. We note that equation~\ref{eq:g_update} is the gain estimator associated with \textsc{StEfCal}.

Equation~\ref{eq:g_update} and~\ref{eq:y_update} were obtained by \citet{Marthi2014} by taking the derivative of the objective function $\Lambda$ relative to the elements of $\bg$ and $\by$, setting the intermediate results to zero and then solving for the unknown parameters (i.e. using the gradient descent algorithm). 
We note that their derivation is less general. The LM algorithm has better convergence properties than gradient descent and 
encompasses the gradient descent algorithm as a special case. In Appendix~\ref{sec:red_stef_ADI} we show that equation~\ref{eq:g_update} and~\ref{eq:y_update} can also be derived using the ADI method.
For this reason, we refer to the implementation of the pseudo-LM calibration scheme derived above, i.e. equation~\ref{eq:g_update} and~\ref{eq:y_update}, as redundant \textsc{StEfCal} throughout the rest of the paper.
Interestingly, \citet{Marthi2014} were not the first to make use of the ADI method to perform redundant calibration, a slower alternative ADI based calibration algorithm is presented in \citet{Wijnholds2012}.

The choice of the $\rho$ parameter is somewhat unconstrained. In this paper we chose $\rho$ by adopting the same strategy that is used by \textsc{StEfCal} and \citet{Marthi2014},
i.e. we chose $\rho$ to be equal to a $\frac{1}{3}$ ($\lambda = 2$). We also carried out simulations to validate this choice.

We generated a sky model that comprised of 100 flat spectrum sources distributed over a 3$^{\circ}$ by 3$^{\circ}$ sky 
patch. The flux density of each source was drawn from a power law distribution with a slope of 2 and the source position was drawn from a uniform distribution. 
We also made use of multiple fictitious telescope layouts each one having a hexagonal geometry (see the left upper image of Figure~\ref{fig:geometry_function} for an example layout). The largest hexagonal array that we used has 217 antennas, with a minimum and maximum baseline of 20~m and 320~m respectively.

We corrupted visibilities by applying gain errors (Table~\ref{tab:gain_parm}) and calibrated the corrupted visibilities using redundant \textsc{StEfCal}. Solutions were independently derived for each time-step and channel for five realizations. 

\begin{table*}
\centering
\caption{The gain error models used in this paper. We have used the symbol $x$ here as a proxy as it can either refer to time-slots or channels. We either
performed our simulations over multiple time-slots and one frequency channel or one timeslot and multiple frequency channels (see Table~\ref{tab:ch_parm}). Moreover, $c$ in the left most column denotes the speed of light.
We adopted a sinusoidal error model, similar to the sinusoidal error models used within \textsc{MeqTrees} \citep{Noordam2010}, as well as a phase slope across frequency which mimics a real case of physical delays between different antennas.
In the first model, we model a gain error with amplitude around one and an additional phase error. In 
the second model we chose $A$ in such a way that we do not unintentionally produce nonsensical gain values, i.e. a zero gain value. For the third model 
the value of $\tau$ was chosen so that the amount of phase wraps that occur across the observing band is restricted to a realistic number.}
\begin{tabular}{|c c c c|} 
\hline
Number tag & 1 & 2 & 3\\
Model & Sinusoidal: amplitude and phase & Sinusoidal: real and imaginary parts & Linear phase slope \\ [0.5ex] 
\hline\hline
Function & $(A+1)e^{jP}$ & $A\cos(2\pi fx+B)+1.5A+jC\sin(2\pi fx+D)$ & $e^{jP}$ \\ 
\hline
Parameters & $A=a\cos(2\pi fx +b)$  & $f=5$ & $P=\tau x$ \\
 & $P =c \cos(2\pi fx +d)$ & $A,C\sim U[0.5,10]$ & $\tau = \frac{l}{c}$ \\
 & $f=5$ & $B,D\sim U[0,2\pi]$ &  $l\sim U[5,50]$ (m)\\
 & $a\sim U[0.8,0.9]$ &  & \\ 
 & $c\sim U[0.5,5]$ &  &  \\ 
 & $b,d\sim U[0,2\pi]$ &  &  \\ 
\hline
\end{tabular}
\label{tab:gain_parm}
\end{table*}

\begin{table}
\centering
\caption{We generated results using two main setups. We either used one frequency channel and multiple time-slots, or one time-slot and multiple frequency channels. The most important parameters used in realizing these two major setups are presented here. 
We chose the observational frequency band of setup 1 to coincide with the HERA array. To broaden the scope of our 
analysis we chose the observational frequency of our second setup to be equal to 1.4~GHz, which is a typical observing frequency of the Westerbork Synthesis Radio Telescope.}
\begin{tabular}{|c c c|} 
\hline
 & Setup 1 & Setup 2\\
\hline
\hline
 Num. channels & 1024 & 1\\
$\nu$-range & 100-200 MHz & 1.4 GHz\\
Num. timeslots & 1 & 50\\
\hline
\end{tabular}
\label{tab:ch_parm}
\end{table}

Note that our simulations are almost ideal as we did not include a primary beam response, nor did we incorporate time and frequency smearing into our simulation.

We also did not explicitly define our noise in terms of the integration time, channel bandwidth and $T_{\textrm{sys}}$, we instead
follow the same approach described in \citep{Liu2010,Marthi2014} and make use of the definition of SNR to introduce noise into our visibilities. 
We use the following definition of 
\citep[SNR, ][]{Liu2010,Marthi2014}:  
\begin{equation}
\label{eq:SNR}
\textrm{SNR} = 10\log \left (\frac{<\bv\odot\conj{\bv}>_{\nu,t,pq}}{<\bn\odot\conj{\bn}>_{\nu,t,pq}} \right ), 
\end{equation}
where $<\mathbf{x}>_{\nu,t,pq}$ denotes averaging over frequency, time and baseline. It should be pointed out, however, that by producing visibilities with different SNR values you are 
effectively either changing your integration time or your channel bandwidth or both (assuming a single $T_{\textrm{sys}}$ value for your instrument). 

\begin{figure}
\includegraphics[width=0.47\textwidth]{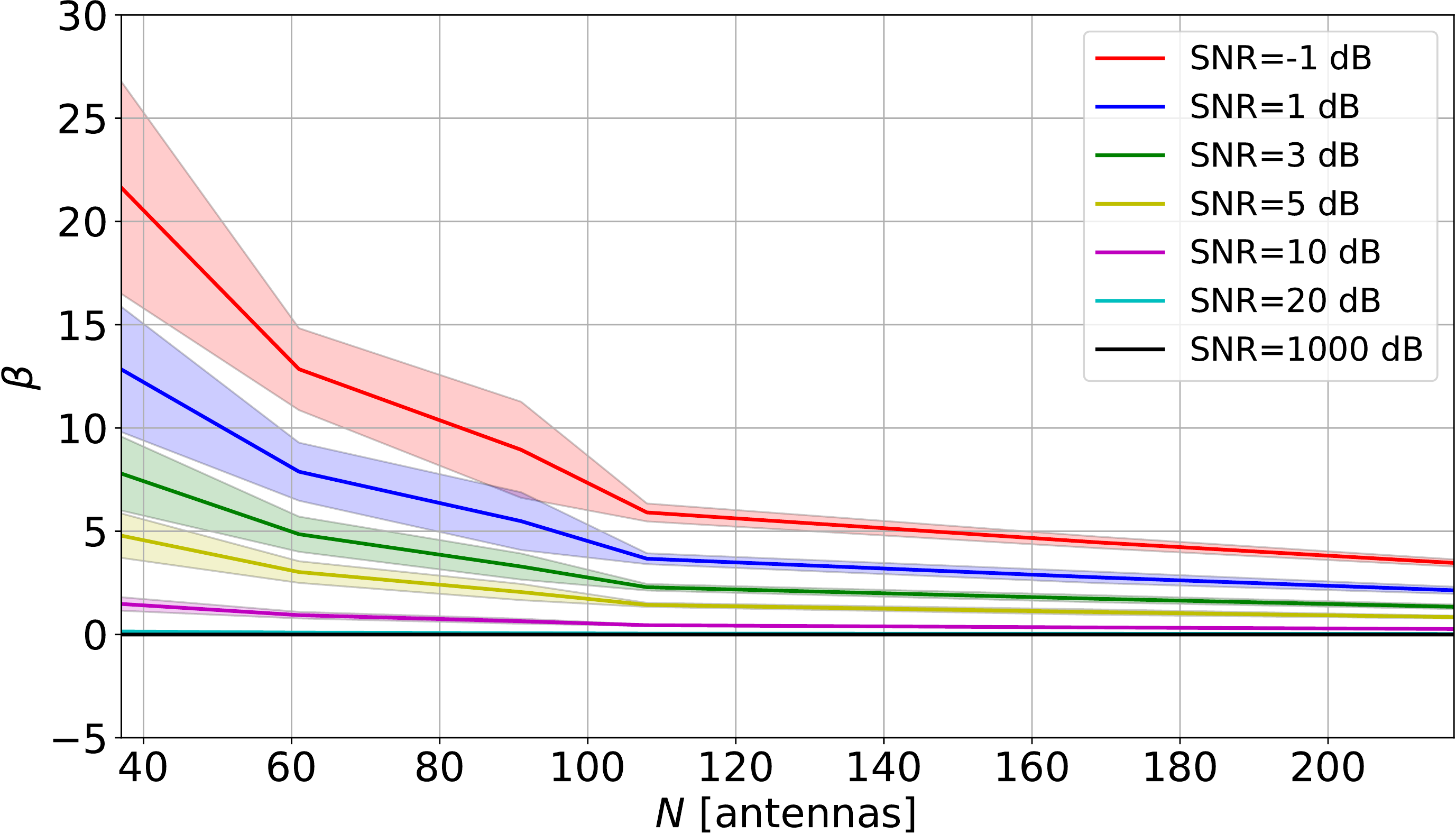} 
\caption{We plot the percentage error $\beta$ between the simulated visibilities and the visibilities solved for by redundant \textsc{StEfCal} for different SNR values as a function of the number of antennas ($N$) in the array.}
\label{fig:prec_error}
\end{figure}

Figure~\ref{fig:prec_error} shows the simulation results as a function of SNR and number of antennas. The accuracy of our solutions is quantified through the percentage error
\begin{equation}
\label{eq:beta}
\beta = \frac{\|\bv - \widehat{\bv}\|_F^2}{\|\bv\|_F^2},
\end{equation}
where $\widehat{\bv}$ is the redundant \textsc{StEfCal} parameter estimate.

The error magnitude follows the expected behaviour, i.e., it decreases as a function of SNR and number of antennas $N$. Interestingly, it reduces to a few percent when $N > 120$ for essentially any choice of SNR. 

\section{Preconditioned Conjugate Gradient Method}
\label{sec:pcg}
\citet{Liu2010} suggested that the execution speed of redundant calibration could be reduced using the 
conjugate gradient method \citep{Hestenes1952}, which would be computationally advantageous, since the Hessian matrix associated 
with redundant calibration (see Figure~\ref{fig:hessian}) is sparse \citep{Reid1971}. In this section we study the computational complexity
of the conjugate gradient (CG) method when it is used to invert the modified Hessian matrix (see equation~\ref{eq:LM_update}), 
in particular when preconditioning is used (i.e. the PCG method). Interestingly, empirical tests suggest that the unmodified Hessian itself is singular. It is therefore important to mention, that the CG method 
can pseudo invert the unmodified Hessian as well, i.e. the CG method can be directly applied to equation~\ref{eq:GN_update}, because the Hessian is a positive semi-definite Hermitian matrix and the vector $\bJ^H\breve{\br}$ is an element of its column range \citep{Lu2015}.


The computational complexity of the CG method is  
\begin{equation}
\label{eq:cg_bound}
O(\sqrt{\kappa}m), 
\end{equation}
where $m$ denotes the number of non-zero entries in and $\kappa$ denotes the spectral 
condition number of the matrix which is to be inverted. The spectral condition number $\kappa$ of the matrix $\bA$ is defined as:
\begin{equation}
\label{eq:kappa}
\kappa(\bA) = \frac{\iota_{\textrm{max}}}{\iota_{\textrm{min}}}, 
\end{equation}
where $\iota_{\textrm{max}}$ and $\iota_{\textrm{min}}$ denote the largest and the smallest eigenvalue of $\bA$ respectively.

Preconditioning is a technique used to improve the spectral condition number of a matrix. Let us consider a generic system of linear equations 
\begin{equation}
\label{eq:prec}
\bA\bx = \bb,
\end{equation}
and a positive-definite Hermitian matrix $\bM$ so that 
\begin{equation}
\label{eq:prec}
\bM^{-1}\bA\bx = \bM^{-1}\bb.
\end{equation}

The matrix $\bM$ is a good preconditioner if:
\begin{equation}
\label{eq:kappa}
\kappa(\bM^{-1}\bA) \ll \kappa(\bA),
\end{equation}
i.e. if it lowers lowers the condition number of a matrix.
If $\bA$ is a nearly diagonal matrix, the Jacobian preconditioner is a natural choice of $\bM$ and can be computed as:
\begin{equation}
\bM = \bA\odot\bI. 
\end{equation}


In order to quantify the effectiveness of the CG method in redundant calibration, we investigate the spectral condition number and the sparsity of the modified Hessian $\bmH$ (i.e. $\lambda = 2$).

\begin{figure*}
\centering
\subfigure[$\kappa$]{\includegraphics[width=0.47\textwidth]{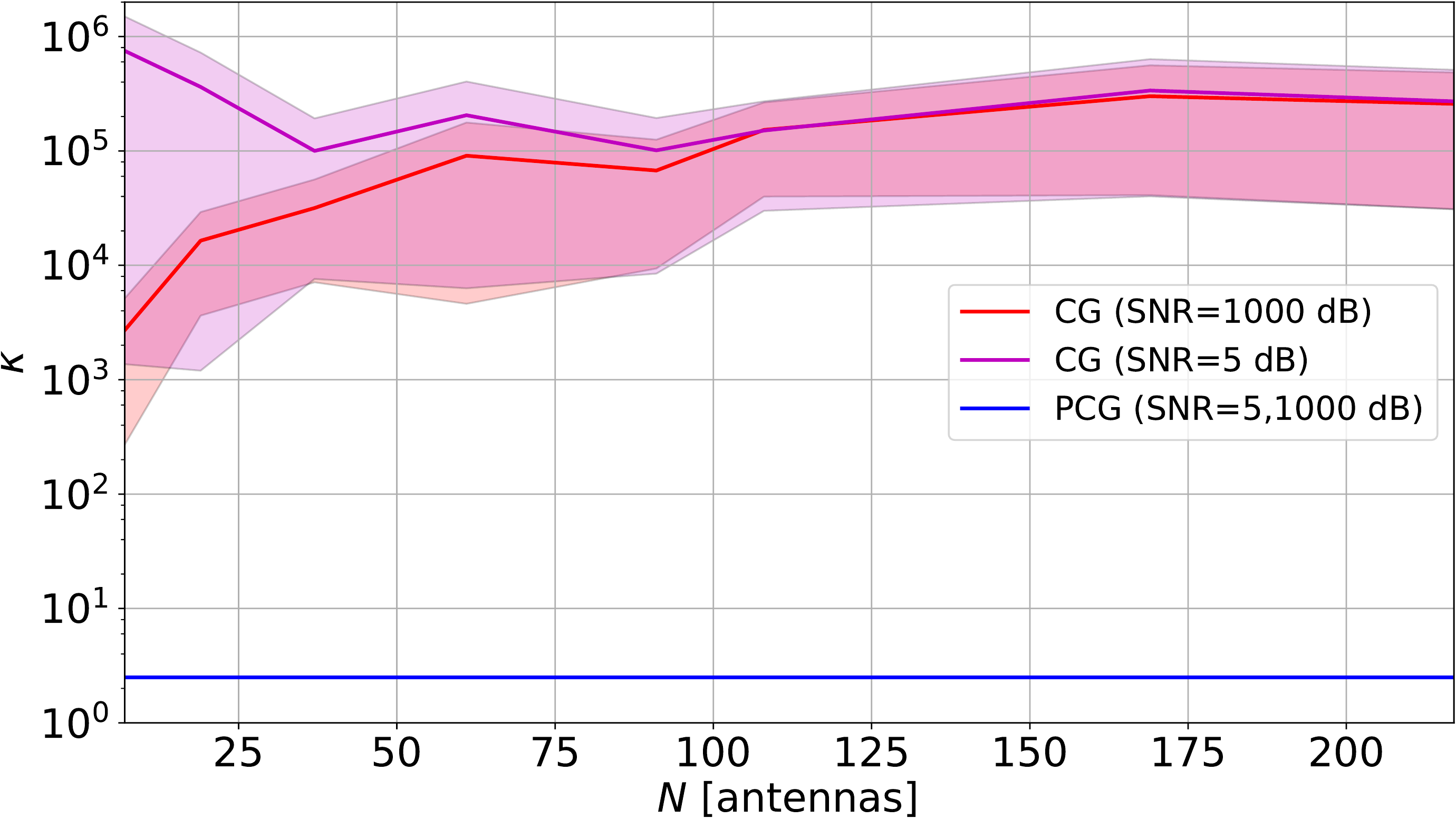}\label{fig:kappa}}
\subfigure[Iterations required before and after Preconditioning]{\includegraphics[width=0.47\textwidth]{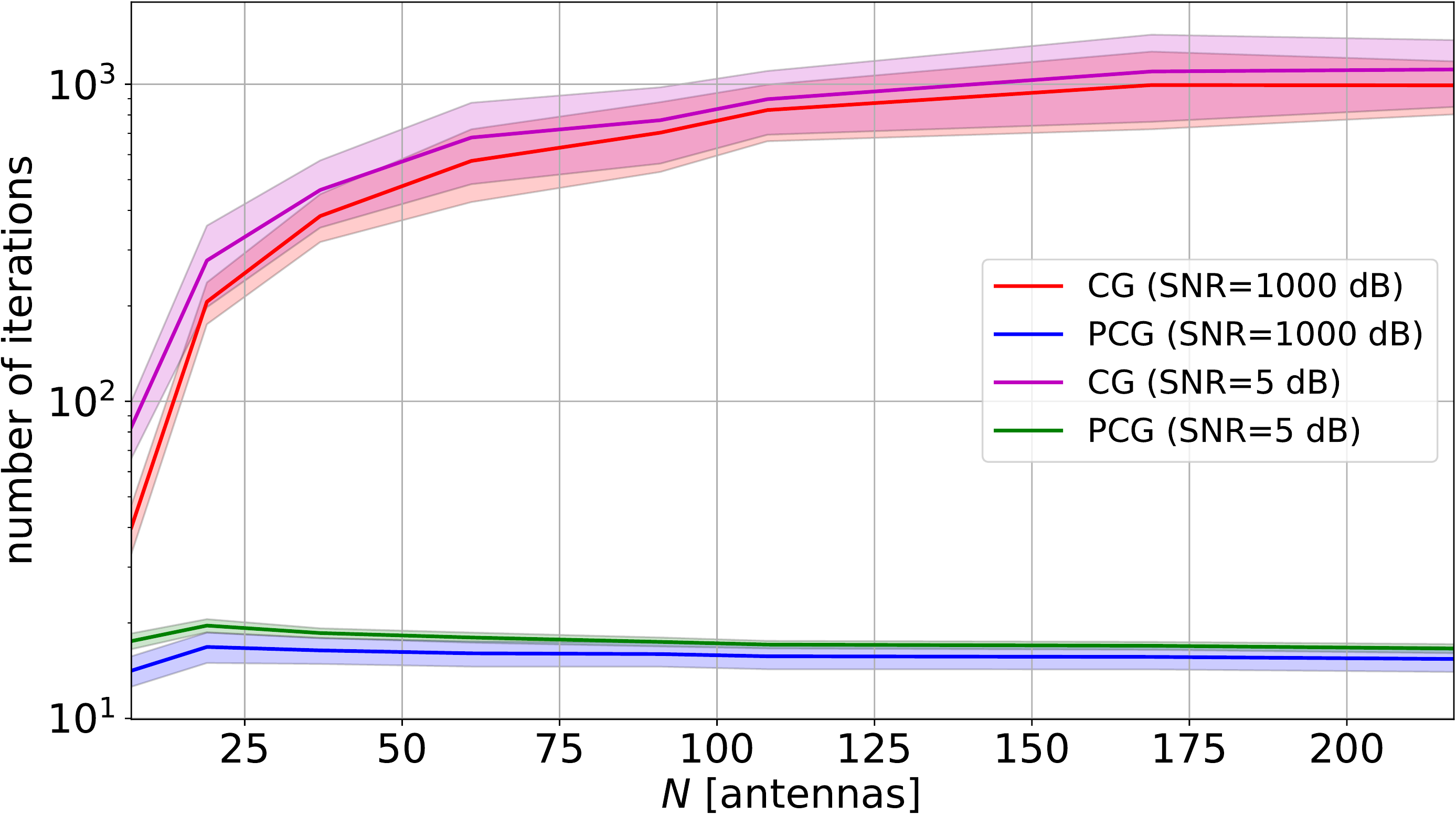}\label{fig:cg_itr}}
\caption{The graphs presented here were generated using the simulations discussed in Section~\ref{sec:red_wirtinger}. Left: spectral condition number $\kappa$ of the modified Hessian $\bmH$ as a function of $N$, before (magenta and red curves) and after (blue curve) preconditioning, at different SNR values. Right: number of major iterations required by the conjugate 
gradient method to invert $\bmH$ as a function of the number of antennas $N$ in the array, before (magenta and red curves) and after preconditioning (blue and green curves), at different SNR values. Both plots show that the Jacobian preconditioner definitely speeds up the conjugate gradient method. \label{fig:kappa_itr}} 
\end{figure*}

We generated simulated (i.e. corrupted) visibilities according to models described in Table~\ref{tab:gain_parm}. 
%
We used the complex LM-algorithm described in Section~\ref{sec:red_wirtinger} to calibrate the corrupted visibilities. To invert the modified
Hessian we used the CG method, with and without a Jacobian preconditioner. 
Figure~\ref{fig:kappa} shows us that preconditioning reduces the condition number of $\bmH$ to a small constant value and therefore effectively eliminates
it from equation~\ref{eq:cg_bound}, i.e. equation~\ref{eq:cg_bound} reduces to
\begin{equation}
\label{eq:cg_bound2}
O(m). 
\end{equation}
This result is confirmed by Fig.~\ref{fig:cg_itr}. Fig.~\ref{fig:cg_itr} 
shows us that the number of major iterations needed by the PCG method to invert 
$\bmH$ is independent of the number of antennas in the array and that it is much less than the 
dimension of $\bmH$. 

Let us now shift our attention towards the remaining factor $m$. To get a useful idea of the computational complexity of CG, we must relate $m$ and $P$. This can be achieved by utilizing the modified Hessian's measure of sparsity $\gamma$:   
\begin{equation}
 \gamma = \left (1 - \frac{m}{P^2} \right ),
\end{equation}
where $m$ is the number of non-zero entries in $\bmH$  and $P^2$ is total number of matrix elements. 

For a regular east-west geometric configuration the sparsity and the asymptotic sparsity of $\bmH$ can be derived analytically:
\begin{align}
\gamma &= \frac{5N^2-7N+3}{8N^2-8N+2} & \gamma_{\infty} &= \lim_{N\rightarrow \infty}\gamma = \frac{5}{8} \label{eq:gamma}. 
\end{align}

\begin{figure*}
\centering
\subfigure[$\gamma$]{\includegraphics[width=0.47\textwidth]{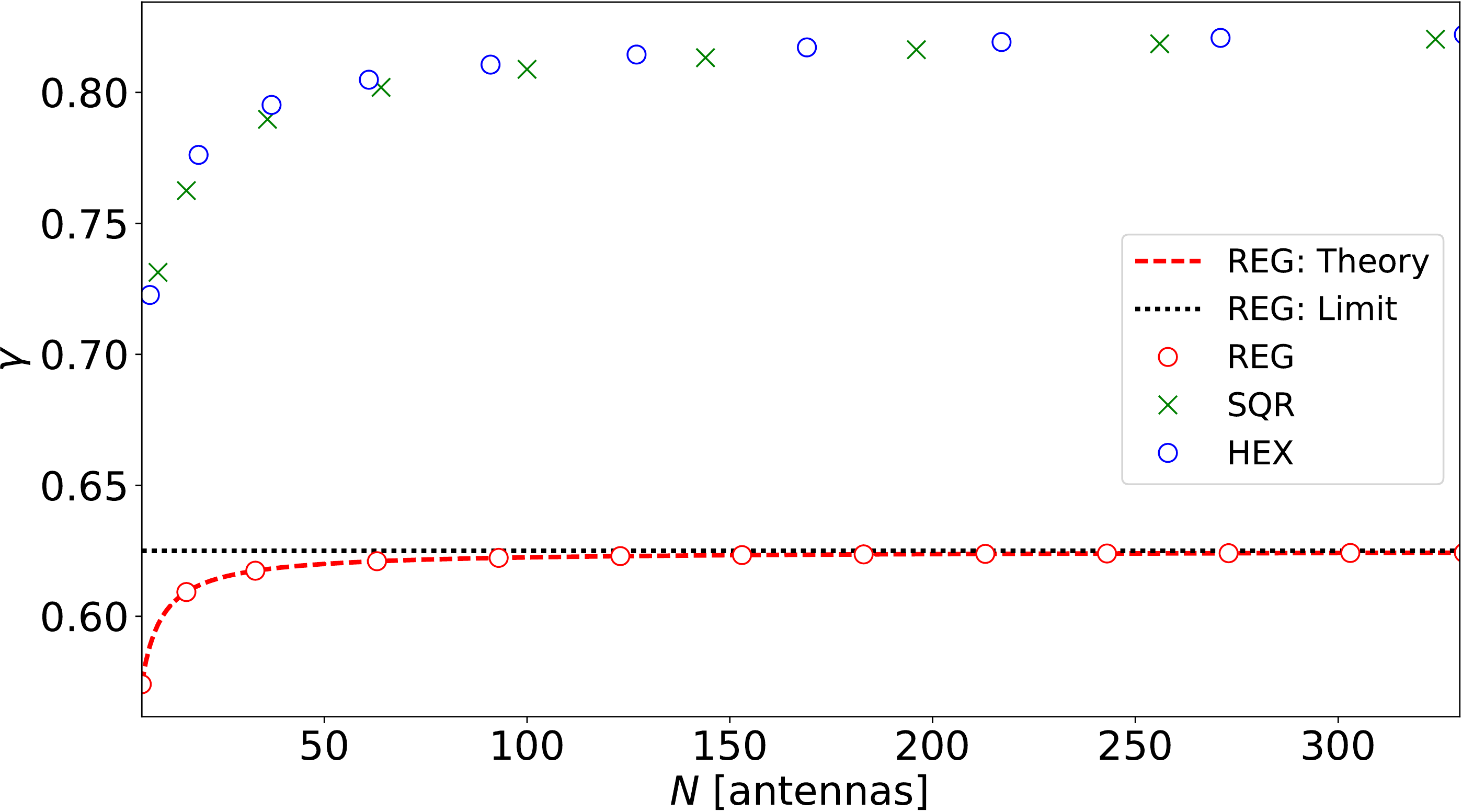}\label{fig:gamma}}
\subfigure[$c$]{\includegraphics[width=0.46\textwidth]{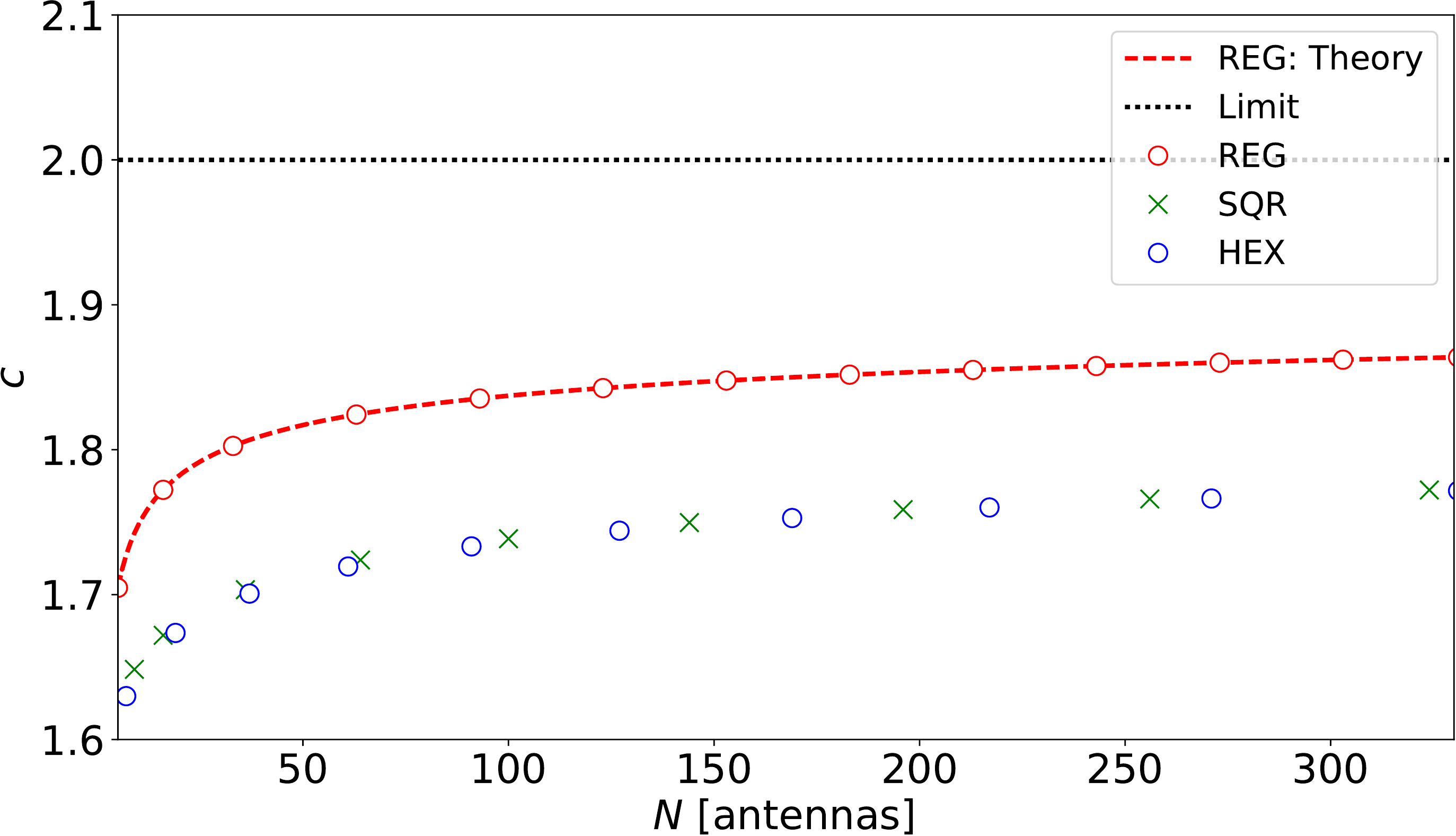}\label{fig:c}}
\caption{Left: the sparsity ratio $\gamma$ of the modified Hessian $\bmH$ as a function of the number of antennas $N$ for an hexagonal (blue circles), square (green crosses) and regular east-west (red circles) array geometry. The red--dashed and black--dotted lines show the analytical expression for $\gamma$ and its limit for the regular east-west grid case (see text for details). Right: the order of the computational cost $c$ for inverting $\bmH$ as a function of $N$ for different array geometries (same colour scheme as in the left panel). The red--dashed and black--dottted lines are the analytical expression of $c$ and its limit in the east-west regular grid case.
\label{fig:sparsity}} 
\end{figure*}

For more complicated array layouts, however, there is no straightforward analytical solution and we empirically determined the sparsity ratios for three different geometric layouts as a function of the number antennas in the array (see Figure~\ref{fig:gamma}). 

It now follows that 
\begin{equation}
P^{c} = m = (1 - \gamma)P^2,
\end{equation}
which leads to
\begin{align}
c &= \log_{P}(1 - \gamma) + 2 & c_{\infty} &= \lim_{N\rightarrow \infty} c = 2. \label{eq:c}
\end{align}
The computational complexity is therefore asymptotically bounded by $O(P^2)$, although it converges very slowly to its asymptotic value, and in general is equal to $O(P^{c})$ (with $c<2$). In the case of an hexagonal geometric layout with $N < 200$ we have that $c \sim 1.7$ (see Figure~\ref{fig:c}).

\begin{figure*}
\centering
\subfigure[Number of LM iterations required by redundant \textsc{StEfCal} and PCG.]{\includegraphics[width=0.45\textwidth]{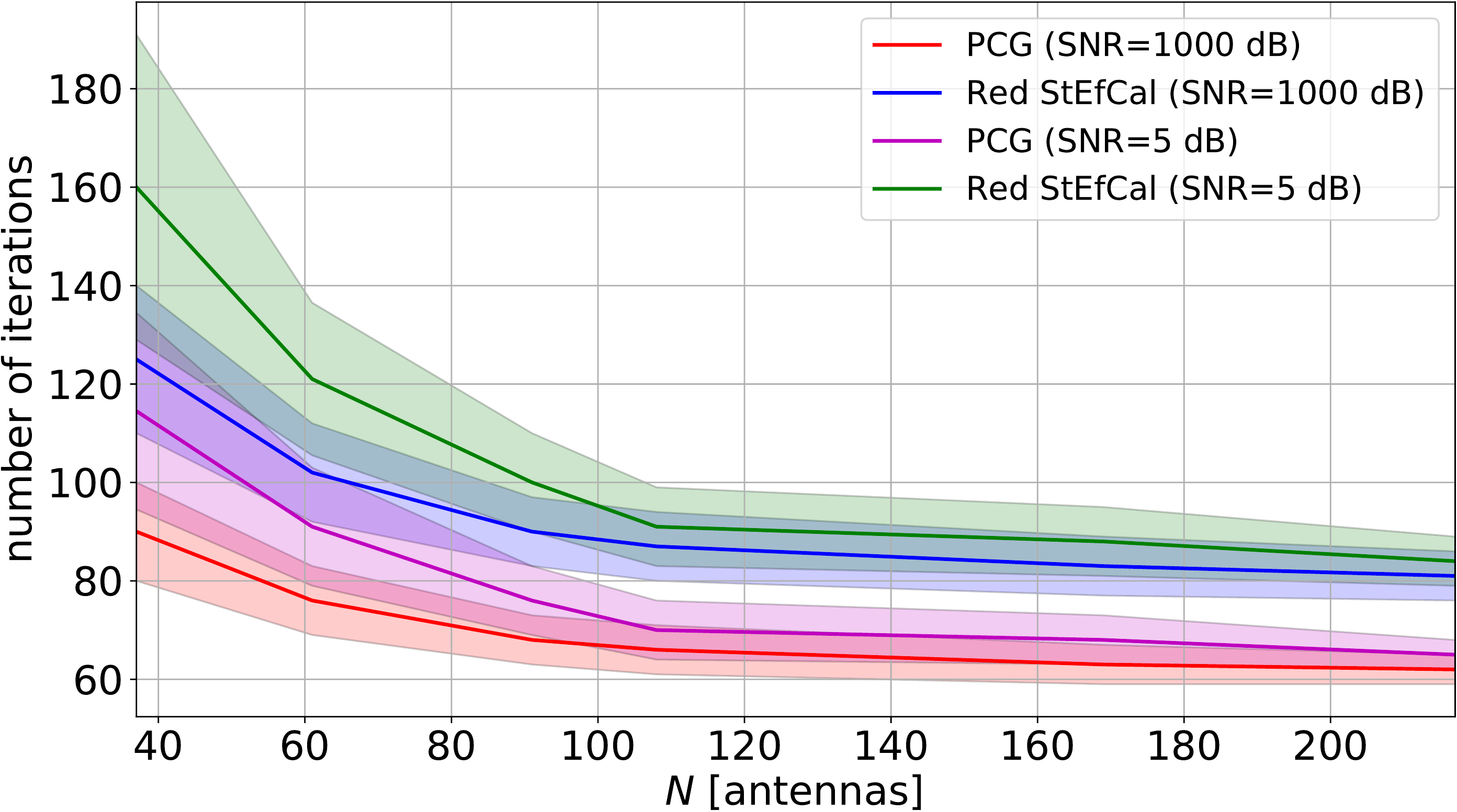}\label{fig:outerloop}}
\subfigure[Difference.]{\includegraphics[width=0.45\textwidth]{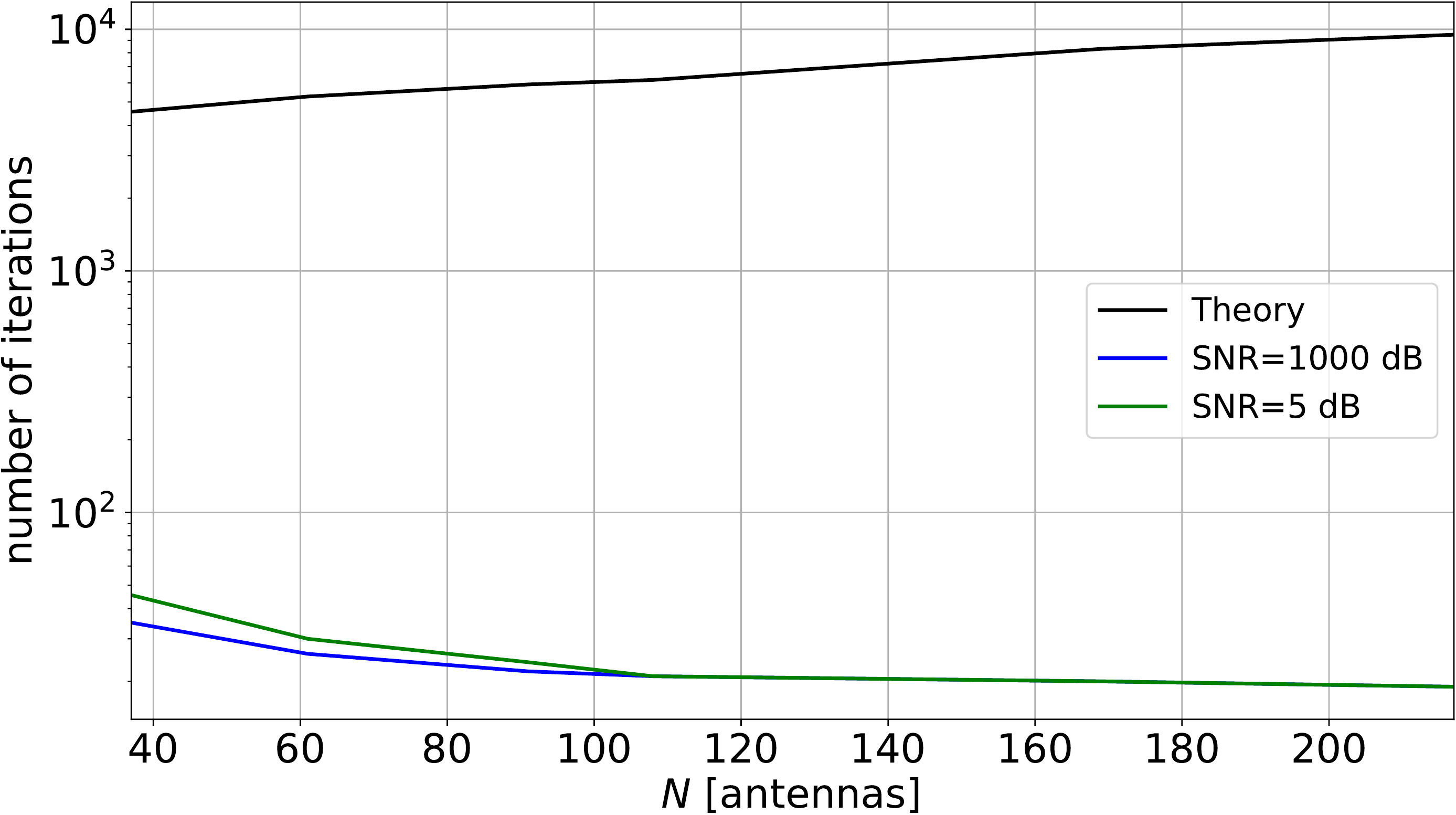}\label{fig:diff}}
\caption{The graphs presented here were generated using the simulations discussed in Section~\ref{sec:red_wirtinger}. Left: number of LM iterations required by redundant \textsc{StEfCal} (green and blue curves) and the PCG (magenta and red curves) methods to converge to a parameter error tolerance of $10^{-6}$ whilst using different SNR values as a function of the number of antennas $N$ in the array. Right: average amount of LM iterations (difference between the redundant \textsc{StEfCal} and PCG curves in the left panel) saved (green and blue curve) by computing the full-inverse with the PCG method whilst using different SNR values. The black curve is $\Delta k$, the theoretical number 
of redundant \textsc{StEfCal} iterations that are needed in excess of the number of LM iterations used for the LM algorithm to 
outperform redundant \textsc{StEfCal}. For the black curve we assumed $c=1.7$ and that $k$ could be approximated with the 
the magenta curve plotted in the left panel.
\label{fig:out_diff}} 
\end{figure*}

We are now finally able to compare the computational complexity of redundant \textsc{StEfCal} and the PCG method. Redundant \textsc{StEfCal} is computationally inexpensive as it just needs to invert a diagonal matrix, however, the PCG inversion is accurate and, therefore, may require fewer iterations, ultimately leading to a faster convergence. 
We computed the theoretical number of redundant \textsc{StEfCal} iterations $\Delta k$ in excess of the number of LM (implementing PCG) iterations required for LM to outperform redundant \textsc{StEfCal}
\begin{align}
\label{eq:k} 
 (k+\Delta k)P >& \, kP^c, & \Delta k >& \, k(P^{c-1}-1).
\end{align}
We then compared $\Delta k$ with the empirically obtained average excess number of iterations. The results are displayed in Figure~\ref{fig:out_diff} which shows that redundant \textsc{StEfCal} outperforms the PCG method. We note that, in this comparison, we have only taken into account the cost of inverting $\bmH$ and ignored the cost of preconditioning. 
%

\section{Conclusion}
\label{sec:conclusions}
In this paper we have formulated the calibration of redundant interferometric arrays using the complex optimization formalism 
\citep{Smirnov2015}. We derived the associated complex Jacobian and the GN and LM parameter update steps.
We also showed that the LM parameter update step can be simplified to obtain an ADI type of algorithm \citep[][]{Salvini2014,Marthi2014}.
Our code implementation of this algorithm (redundant \textsc{StEfCal}) is publicly available at 
\url{https://github.com/Trienko/heracommissioning/blob/master/code/stef.py}. 
We note that, in its current implementation, redundant \textsc{StEfCal} does not solve for the degeneracies inherent to redundant calibration  \citep{Zheng2014,Kurien2016,dillon2017} which will be the subject of future work. Compared to current redundant calibration algorithms, redundant \textsc{StEfCal} is more robust to inital conditions and allows for easier parallelization. 

We investigated the computational cost of redundant \textsc{StEfCal} and compared it with the performance of the PCG method (as suggested by \cite{Liu2010}).
We found that, although the PCG method greatly improves the speed of redundant calibration, it still significantly underperforms when compared to redundant \textsc{StEfCal}.

The characteristics of redundant \textsc{StEfCal} make it an appealing calibration algorithm for large redundant arrays like HERA \citep{deboer2017}, CHIME \citep{Bandura2014}, HIRAX \citep{Newburgh2016} or even hybrid arrays like the MWA \citep{Tingay2013} in its updated phase.

\section{Future Outlook}
We plan to apply redundant \textsc{StEfCal} to HERA data in the near future. The reason for applying redundant \textsc{StEfCal} to real data is two-fold. First, we wish to validate the theoretical computational
complexity of redundant \textsc{StEfCal} (derived earlier in the paper). Second, we would like to test whether it could be used to calibrate HERA in near-realtime. 
We are also interested in parallelizing our current implementation and to see how much can be gained by doing so. We would also like to 
conduct a perturbation analyses, similar to the one conducted by \cite{Liu2010}, to estimate the error which is introduced into our 
estimated gains and visibilities when the array contains baselines which are not perfectly redundant. We are also interested in quantifying 
the error which is introduced into the estimated gains and visibilities because of differing primary beam patterns between array elements \citep{Noorishad2012}. 

\section*{Acknowledgements}
This work is based upon research supported by the South African Research Chairs Initiative of the Department of Science and Technology and National Research Foundation. GB acknowledges support from the Royal Society and the Newton Fund under grant NA150184. This work is based on the research supported in part by the National Research Foundation of South Africa (grant No. 103424).

We would like to thank the anonymous reviewer for their useful comments, which greatly improved the quality of the paper. 

\bibliographystyle{mn2e}
\bibliography{paper}

\appendix

\section{Derivation of redundant Wirtinger calibration}
\label{sec:analytic}
If we apply the definition in equation~\ref{eq:Jacobian} to equation~\ref{eq:least_squares_complex} we obtain the following analytic result:
\begin{equation}
\label{eq:Jacobian_red}
\bJ = \begin{bmatrix}
       \bM & \bN\\
       \conj{\bN} & \conj{\bM}
      \end{bmatrix},
\end{equation}
where
\begin{equation}
\bM =\begin{bmatrix}
      \bO & \bP
     \end{bmatrix},
\end{equation}
and 
\begin{equation}
\bN = \begin{bmatrix}
       \bQ & \bzero
      \end{bmatrix}.
\end{equation}
Moreover,
\begin{equation}
\left [ \bO  \right ]_{\alpha_{pq},j} = \begin{cases}
                                         y_{\phi_{pq}}\conj{g_q} & \textrm{if}~p=j\\
                                         0  & \textrm{otherwise} 
                                        \end{cases},
\end{equation}

\begin{equation}
\left [ \bP  \right ]_{\alpha_{pq},j} = \begin{cases}
                                         g_p\conj{g_q} & \textrm{if}~\phi_{pq}=j\\
                                         0  & \textrm{otherwise} 
                                        \end{cases}
\end{equation}
and
\begin{equation}
\left [ \bQ  \right ]_{\alpha_{pq},j} = \begin{cases}
                                         g_py_{\phi_{pq}} & \textrm{if}~q=j\\
                                         0  & \textrm{otherwise} 
                                        \end{cases}
\end{equation}

We use $\bzero$ to denote an all zero matrix. It is now trivial to compute the Hessian $\bH$ by using equation~\ref{eq:Jacobian_red}. If we substitute equation~\ref{eq:Jacobian_red} into $\bJ^H\bJ$
we obtain 
\begin{equation}
\label{eq:red_H}
\bH = \bJ^H\bJ = 
\begin{bmatrix}
\bA & \bB\\
\conj{\bB} & \conj{\bA}
\end{bmatrix},
\end{equation}
where

\begin{align}
\bA &= \begin{bmatrix} \bC & \bD\\ \bD^H & \bE \end{bmatrix}, & \bB &= \begin{bmatrix} \bF & \bG\\ \bG^T & \bzero \end{bmatrix},
\end{align}

\begin{equation}
[\bC]_{ij} = 
\begin{cases}
 \sum_{k \neq i} \left | g_k \right |^2 \left | y_{\zeta_{ik}} \right |^2 & \textrm{if} ~ i=j\\
 0 & \textrm{otherwise}
\end{cases},
\end{equation}
\begin{equation}
\label{eq:D_mat}
[\bD]_{ij} = 
\begin{cases}
 g_i \conj{y}_j  \left | g_{\psi_{ij}} \right |^2  & \textrm{if} ~ \psi_{ij}\neq0\\
 0 & \textrm{otherwise}
\end{cases},
\end{equation}

\begin{equation}
[\bE]_{ij} = 
\begin{cases}
 \sum_{rs \in \mathcal{RS}_i} \left | g_r \right |^2 \left | g_s \right |^2  & \textrm{if} ~ i=j\\
 0 & \textrm{otherwise}
\end{cases},
\end{equation}
\begin{equation}
[\bF]_{ij} = 
\begin{cases}
 g_i g_j  \left | y_{\zeta_{ij}} \right |^2  & \textrm{if} ~ i \neq j\\
 0 & \textrm{otherwise}
\end{cases},
\end{equation}
and
\begin{equation}
[\boldsymbol{G}]_{ij} = 
\begin{cases}
 g_i y_j  \left | g_{\xi_{ij}} \right |^2  & \textrm{if} ~ \xi_{ij}\neq0\\
 0 & \textrm{otherwise}
\end{cases}.
\end{equation}
Moreover, 
\begin{equation}
\label{eq:RS}
\mathcal{RS}_i = \left\{rs\in\mathbb{N}^2|(\phi_{rs} = i) \right\},
\end{equation}
\begin{equation}
\xi_{ij} = 
\begin{cases}
p~\textrm{if}~\exists! ~ p \in \mathbb{N} ~ s.t. ~(\phi_{pi} = j)\\
0~\textrm{otherwise}
\end{cases},
\end{equation}
and
\begin{equation}
\psi_{ij} = 
\begin{cases}
q~\textrm{if}~\exists! ~ q \in \mathbb{N} ~ s.t. ~(\phi_{iq} = j)\\
0~\textrm{otherwise}
\end{cases}.
\end{equation}

Furthermore, substituting equation~\ref{eq:Jacobian_red} into $\bJ^H\breve{\br}$ results in
\begin{equation}
\bJ^H\breve{\br} = \begin{bmatrix}
                   \ba \\
                   \bb \\
                   \conj{\ba}\\
                   \conj{\bb}
                   \end{bmatrix},
\end{equation}
where
\begin{align}
\label{eq:ab}
\left [ \ba \right ]_i &= \sum_{k\neq i} g_k \widetilde{y}_{ik}r_{ik},  & \left [ \bb \right ]_i &= \sum_{rs\in\mathcal{RS}_i}\conj{g}_r g_s r_{rs},
\end{align}
and
\begin{equation}
\label{eq:y_tilde}
\widetilde{y}_{ik} = 
\begin{cases}
\conj{y}_{\zeta_{ik}} & \textrm{if}~k > i\\
y_{\zeta_{ik}} & \textrm{otherwise}
\end{cases}.
\end{equation}
Additionally, $\ba$ and $\bb$ are both column vectors. The lengths of $\ba$ and $\bb$ are $N$ and $L$ respectively.
The dimensions of the matrices we defined in this section are presented in Table~\ref{tab:matrix_dimensions}.

Furthermore,
\begin{align}
\label{eq:identities}
\frac{1}{3}\bJ\breve{\bz} &= \breve{\bv}, & \bJ^H\breve{\bv} &= (\bI\odot\bH)\breve{\bz} 
\end{align}
The above identities can be trivially established by mechanically showing that the left hand side of each expression in equation~\ref{eq:identities} is equal to its right hand side.

\begin{table}
\centering
\caption{The dimensions of the matrices defined in Appendix~\ref{sec:analytic}.}
\begin{tabular}{|c c|} 
\hline
Matrix & Dimension\\
\hline
\hline
$\bJ$ & $2B \times P$ \\
$\bM$ & $B \times (N+L)$ \\
$\bN$ & $B \times (N+L)$ \\
$\bO$ & $B \times N$ \\
$\bP$ & $B \times L$ \\
$\bQ$ & $B \times N$ \\
$\bzero$ & $B \times L$ \\
\hline
\hline
$\bH$ & $P\times P$\\
$\bA$ & $(N+L)\times (N+L)$\\
$\bB$ & $(N+L)\times (N+L)$\\
$\bC$ & $N \times N$\\
$\bD$ & $N \times L$\\
$\bE$ & $L \times L$\\
$\bF$ & $N \times N$\\
$\bG$ & $N \times L$\\
$\bzero$ & $L \times L$\\
\hline
\end{tabular}
\label{tab:matrix_dimensions}
\end{table}

\section{ADI}
\label{sec:red_stef_ADI}
The basic skymodel-based \textsc{StEfCal} update step is equal to  the leftmost term in equation~\ref{eq:g_update} (barring $\rho$) \citep{Salvini2014}.
Assume without any loss of generality that the array is in an east-west regular grid. Furthermore, assume that $\boldsymbol{d}$ (see equation~\ref{eq:vis_linear_definition}) has been re-ordered 
and that the result of this re-ordering is the following
\begin{equation}
\widetilde{\boldsymbol{d}} = \left[d_{12},\cdots,d_{N-1,N},d_{13},\cdots,d_{N-2,N},\cdots,d_{1N}\right]^T .
\end{equation}
The vector $\widetilde{\bn}$ should be interpreted in a similar manner.

Equation~\ref{eq:vis_linear_definition} can now be rewritten as 
\begin{equation}
\label{eq:linear_system}
\widetilde{\boldsymbol{d}} = \boldsymbol{J}\boldsymbol{y} + \widetilde{\bn}, 
\end{equation}
if we assume that $\boldsymbol{g}$ and its conjugate are known vectors. In equation~\ref{eq:linear_system},
\begin{equation}
\boldsymbol{J} = 
\begin{bmatrix}
g_1\conj{g}_2 & 0 & \cdots & 0\\
g_2\conj{g}_3 & 0 & \cdots & 0\\
\vdots & 0 & \cdots & 0\\
g_{N-1}\conj{g}_N & 0 & \cdots & 0\\
0 & g_1\conj{g}_3 & \cdots & 0\\
0 & \vdots & \cdots & 0\\
0 & g_{N-2}\conj{g}_N & \cdots & 0\\
0 & 0 & \cdots & 0\\
  & \vdots & \\
0 & 0 & \cdots & g_1\conj{g}_N\\  
\end{bmatrix}.
\end{equation}
We can now estimate $\boldsymbol{y}$ with
\begin{equation}
\label{eq:y_final}
\boldsymbol{y} = (\boldsymbol{J}^H\boldsymbol{J})^{-1}\boldsymbol{J}^H\widetilde{\boldsymbol{d}}, 
\end{equation}
where 
\begin{equation}
\label{eq:RHR}
[\boldsymbol{J}^H\boldsymbol{J}]_{ij} = 
\begin{cases}
\sum_{rs\in\mathcal{RS}_i} |g_r|^2|g_s|^2 &\textrm{if}~i=j\\
0&\textrm{otherwise}
\end{cases},
\end{equation}
and
\begin{equation}
\label{eq:RHd}
[\boldsymbol{J}^H\widetilde{\boldsymbol{d}}]_i = \sum_{rs\in\mathcal{RS}_i} \conj{g}_r g_s d_{rs}. 
\end{equation}
Substituting equation~\ref{eq:RHR} and equation~\ref{eq:RHd} into equation~\ref{eq:y_final} and simplifying the result leads to the leftmost term in equation~\ref{eq:y_update} (if we bar $\rho$ and consider 
only the $i$-th entry of $\by$).
\label{lastpage}
\end{document}